\documentclass[10pt]{article}
\usepackage[dvips]{graphicx}

\usepackage{epsfig}
\usepackage{graphicx}
\usepackage{indentfirst}
\usepackage{amsmath}
\usepackage{amsfonts}
\usepackage{amssymb}

\begin{document}

\begin{center}
{\Large\bf Testing some f(R,T) gravity models from energy conditions\\}

\medskip

 F. G. Alvarenga$^{a}$\footnote{e-mail: f.g.alvarenga@gmail.com},\,\, M. J. S. Houndjo$^{a,b}$\footnote{e-mail:
sthoundjo@yahoo.fr},\,\,A. V. Monwanou$^{b}$\footnote{e-mail: movins2008@yahoo.fr}\,\,and Jean B. Chabi Orou$^{b}$\footnote{e-mail: jean.chabi@imsp-uac.org}\\ 
 $^{a}${\it Departamento de Ci\^{e}ncias Naturais - CEUNES - 
Universidade Federal do Esp\'irito Santo\\
CEP 29933-415 - S\~ao Mateus - ES, Brazil}
\\

 $^{b}${\it Institut de Math\'{e}matiques et de Sciences Physiques (IMSP), 01 BP 613 Porto-Novo, B\'{e}nin} 
\medskip\\
 \end{center}

\begin{abstract}
We consider $f(R, T)$ theory of gravity, where $R$ is the curvature scalar and $T$ the trace of the energy momentum tensor. Attention is attached to the  special case, $f(R, T)= R+2f(T)$ and two expressions are assumed for the function $f(T)$, $\frac{a_1T^n+b_1}{a_2T^n+b_2}$ and $a_3\ln^{q}{(b_3T^m)}$, where $a_1$, $a_2$, $b_1$, $b_2$, $n$, $a_3$, $b_3$, $q$ and $m$ are input parameters. We observe that by adjusting suitably  these input parameters, energy conditions can be satisfied.  Moreover,  an analyse of the perturbations and stabilities of de Sitter solutions and power-law solutions is performed with the use of the two models. The results  show that for some values of the input parameters, for which energy conditions are satisfied, de Sitter solutions and power-law solutions may be stables.
\end{abstract}

Pacs numbers: 04.50.-h, 04.50.Kd, 98.80.-k

\section{Introduction}
It is well known that General Relativity (GR) based on the Einstein-Hilbert action  (without taking into account the dark energy) can not explain the acceleration of the early and late universe. Therefore, GR does not describe precisely gravity and it is quite reasonable to modify it in order to get theories that admit inflation and imitate the dark energy. The first tentative in this way is substituting Einstein-Hilbert term by an arbitrary function of the curvature scalar $R$, this is the so-called f$(R)$ theory of gravity. This theory has been widely studied and interesting results have been found \cite{papierfR1}-\cite{papierfR11}. In the same way, other alternative theory of modified gravity has been introduced, the so-called Gauss-Bonnet gravity, $f(G)$, as a general function of the Gauss-Bonnet invariant term $G$ \cite{gauss1}.   Other combinations of scalars  are also used as the generalised $f(R,G)$ and $f(R,P,Q)$ \cite{other1,other2}, where $P\equiv R_{\mu\nu}R^{\mu\nu}$ and $Q\equiv R_{\mu\nu\sigma\tau}R^{\mu\nu\sigma\tau}$ (here $R_{\mu\nu}$ and $R_{\mu\nu\sigma\tau}$ are the Ricci tensor the Riemann tensor, respectively).\par 
In this present paper, attention is attached to a type of the so-called $f(R, T)$ theory of gravity, where $T$ denotes the trace of the energy momentum tensor. This generalization of $f(R)$ gravity has been made first by Harko et al \cite{odintsovfRT}. The dependence on $T$ can take source from the introduction of exotic imperfect fluid or from quantum effect (conformal anomaly). In this theory, the equations of motion show the presence of an extra-force acting on the test particles, and the motion are generally non-geodesic. This theory also relates  the cosmic acceleration, not only due to the contribution of geometrical terms, but also to the matter contents \cite{odintsovfRT}. Some results have been obtained with  this theory. In  \cite{stephaneseul3}, the cosmological reconstruction of $f(R, T)$ describing transition from matter dominated phase to the late accelerated epoch of the universe is performed. Also in the same way for exploring  cosmological scenarios based on this theory, $f(R,T)$ function has been numerically reproduced according to holographic dark energy \cite{oliver}. Moreover it is shown that dust reproduces $\Lambda CDM$, phantom-non-phantom and the phantom cosmology with $f(R,T)$ theory \cite{ratbay}.     The general technique for performing this reproduction of $\Lambda CDM$ model in FRW\footnote{Friedmann-Robertson-Walker} metric cosmological evolution is widely developed in \cite{other1,other3}. The $f(R, T)$ models that are able to reproduce the fourth known types of future finite-time singularities have been investigated \cite{juliano}. The authors also introduced conformal anomaly as quantum effects near the singularities and observe that it cannot remove the sudden singularity or the type IV one, but, for some values of a input parameter, the big rip and the type III singularity may be avoided from the effects of the conformal anomaly. Moreover, they found as interesting result that, even without  taking into account conformal anomaly, singularities (the Big rip and the type-III) may still be removed  thanks to the presence of the  $T$ contribution of  the $f(R, T)$ theory.\par
Note that singularities appear when energy conditions are violated. Our task in this paper is to check the viability of some models of $f(R,T)$ according to the energy conditions. The energy conditions are formulated by the use of the Raychaudhuri equation for expansion and is based on the attractive character of the gravity. We refer the readers to Refs \cite{harko14}-\cite{ecsetare}, where  energy conditions are widely analysed  for the cosmology settings, in f(R) and f(G) gravities. \par
In this paper, we assume a special form of $f(R, T)$, that is,  $f(R, T)=R+2f(T)$,  the usual Einstein-Hilbert term plus a $T$ dependent function  $f(T)$. Two expressions of $f(T)$,   $\frac{a_1T^n+b_1}{a_2T^n+b_2}$ and $a_3\ln^{q}{(b_3T^m)}$ are investigated, where $a_1$, $a_2$, $b_1$, $b_2$, $n$,  $a_3$, $b_3$, $q$ and $m$ are parameters to be suitably adjusted in order to obtain viable models according to energy conditions. The first expression is similar to that used in \cite{model1,model2}, where, at the place of the trace $T$, the Gauss-Bonnet invariant is used. The main motivation of this choice is explained in the section $4$. The second form is chosen due to its interesting aspect, because its corresponding $f(R, T)$ model avoids itself the appearance of the type-I finite-time future singularity (Big Rip) \cite{juliano}, thanks to the ordinary trace $T$-terms contributions. Thus it is quite reasonable to check for which values of the input parameters  this model can be acceptable as cosmological model.\par 
 In order to reach the acceptable cosmological models, we analyse the perturbations and stabilities of de Sitter solutions and power-laws solutions in the framework of  the special $R+2f(T)$ gravity, by using the two models proposed in this work. We observe that for some values of the input parameters, for both models, the stabilities of de Sitter solutions and power-law solutions are realized and compatibles with some energy conditions and the late time acceleration of the universe.
\par
The paper is outlined as follows. In section $2$, we briefly present the general formalism of the theory, putting out the general equations of motion for a  $f(R,T)=f_1(R)+f_2(T) $  gravity, where $f_1(R)$ and  $f_2(T)$ are respectively function of the curvature scalar and the trace of the energy momentum tensor. The section $3$ is devoted to the general aspects of the energy conditions. The $f(R, T)=R+2f(T)$ gravity is assumed in the section $4$, where the two functions considered for $f(T)$ are studied, putting out the conditions on the input parameters for obtaining some viable models of $f(R, T)$.  The perturbations and stabilities of de Sitter and power-law solutions are investigated in the sections $5$. Discussions and perspectives are presented in the section $6$.
\section{General formalism}
Let us assume the modified gravity replacing the Ricci scalar $R$ in Einstein gravity by an arbitrary function $f(R, T)$, and writing the total action as
\begin{eqnarray}\label{vincent1}
S=\frac{1}{2\kappa^2}\int d^4x \sqrt{-g}\left[f(R, T)+\mathcal{L}\right]\,,
\end{eqnarray}
where $\kappa^2=8\pi G$, $G$ being the gravitational constant and $T=g^{\mu\nu}T_{\mu\nu}$ the trace of the matter energy momentum tensor $T_{\mu\nu}$ which is defined by 
\begin{eqnarray}\label{vincent2}
T_{\mu\nu}=-\frac{2}{\sqrt{-g}}\frac{\delta(\sqrt{-g}\mathcal{L})}{\delta g^{\mu\nu}}\,\,.
\end{eqnarray}
This modified gravity theory  has been considered first in \cite{odintsovfRT} and the equations of motion, using the metric formalism, have been explicitly obtained as
\begin{eqnarray}\label{vincent3}
f_R(R, T)R_{\mu\nu}-\frac{1}{2}f(R, T)g_{\mu\nu}+\left(g_{\mu\nu}\Box - \nabla_{\mu}\nabla_{\nu}\right)f_R(R, T) = \kappa^2 T_{\mu\nu}-f_T(R, T)T_{\mu\nu}-f_T(R, T)\Theta_{\mu\nu}\,,
\end{eqnarray}
where $f_R(R, T)$ and $f_T(R,T)$ denote the derivative of f(R, T) with respect to R and T, respectively, and $\Theta_{\mu\nu}$ is defined  by \cite{odintsovfRT}
\begin{eqnarray}\label{vincent4}
\Theta_{\mu\nu}\equiv g^{\alpha\beta}\frac{\delta T_{\alpha\beta}}{\delta g^{\mu\nu}}= -2T_{\mu\nu}+g_{\mu\nu}\mathcal{L}-2g^{\alpha\beta}\frac{\partial^2 \mathcal{L}}{\partial g^{\mu\nu}\partial g^{\alpha\beta}}\,\,.
\end{eqnarray}
Let us consider the stress-energy tensor of the matter as
\begin{eqnarray}\label{vincent5}
T_{\mu\nu}=(\rho+p)u_{\mu}u_{\nu}-pg_{\mu\nu}\,\,.
\end{eqnarray}
 This expression is a consequence of a statement of the fluid mechanics for which a possible way for treating the scalar invariant $\mathcal{L}$ is to interpret it  as the pressure p of the fluid, i.e.  $\mathcal{L}=-p$ (the negative sign is due to  the signature used here) \cite{odintsovfRT,lagrangianpressure}. Here, $\rho$, $p$ and $u^{\mu}$ are respectively the energy density, the pressure and the four-velocity of a perfect fluid considered as the classical matter content of the universe. Hence, Eq.(\ref{vincent4}) becomes
\begin{eqnarray}\label{vincent6}
\Theta_{\mu\nu}=-2T_{\mu\nu}-pg_{\mu\nu}\,\,.
\end{eqnarray}
From now, we will treat the models of type $f(R, T)= f_1(R)+f_2(T)$ and set $\kappa^2=1$. Then, the equation (\ref{vincent3}) becomes
\begin{eqnarray}\label{vincent7}
f_{1R}(R)R_{\mu\nu}-\frac{1}{2}f_1(R)g_{\mu\nu}+\left( g_{\mu\nu}\Box-\nabla_\mu\nabla_\nu\right) f_{1R}(R)=T_{\mu\nu}+f_{2T}(T)T_{\mu\nu}+\left[ f_{2T}(T)p+\frac{1}{2}f_2(T)\right] g_{\mu\nu}
\end{eqnarray}
\section{Energy conditions}
 The energy conditions are essentially based on the Raychaudhuri equation that describes the behaviour of a congruence of timelike, spacelike or lightlike curves. It is commonly used to study and establish singularities of the spacetime. Considering a congruence of the curves fulfilling the spacetime manifold and also that there are expansion, distortion and relative rotation between the curves of the congruence, then the spacetime manifold described by the congruence is curved. For the purposes of this work we will just consider the timelike and  spacelike curves for which the Raychaudhuri equation reads, respectively \cite{hawking,tahim}
\begin{eqnarray}\label{movins1}
R_{\mu\nu}V^{\mu}V^{\nu}+\frac{1}{3}\theta^2+\sigma_{\mu\nu}\sigma^{\mu\nu}-\omega_{\mu\nu}\omega^{\mu\nu}+\frac{d\theta}{d\tau}=0\,\,,\\
R_{\mu\nu}k^{\mu}k^{\nu}+\frac{1}{2}\theta^2+\sigma_{\mu\nu}\sigma^{\mu\nu}-\omega_{\mu\nu}\omega^{\mu\nu}+\frac{d\theta}{d\lambda}=0\,\,,
\end{eqnarray}
where $\theta$ is the expansion scalar describing the expansion of volume, $\tau$ and $\lambda$ are positive parameters used to describe the curved of the congruence, $\sigma_{\mu\nu}$ the shear tensor which measures the distortion of the volume, $\omega_{\mu\nu}$ the vorticity tensor which measures the rotation of the curves, and $V^{\mu}$ and $k^{\mu}$ are respectively timelike and lightlike  vectors tangent to the curves. In this work, we are interested to the situation for small distortions of the volume, without rotation, in such a way that the quadratic terms in the Raychaudhuri equation may be disregarded (they are like second order corrections). Then, the equation can be integrated given the scalar of expansion as a function of the Ricci tensor:
\begin{eqnarray}
\theta = -\tau R_{\mu\nu}V^{\mu}V^{\nu}= -\lambda R_{\mu\nu}k^{\mu}k^{\nu} \,\,.
\end{eqnarray}
The condition for attractive gravity is $\theta<0$, imposing $R_{\mu\nu}V^{\mu}V^{\nu}\geq 0$ and $R_{\mu\nu}k^{\mu}k^{\nu}\geq 0$. These two conditions are called the strong and null energy conditions, respectively. \par
For equivalence to GR, by just dividing by $f_1(R)$ (different from zero), one can cast Eq. (\ref{vincent7}) in the following form 
\begin{eqnarray}
R_{\mu\nu}-\frac{1}{2}Rg_{\mu\nu}= T_{\mu\nu}^{eff}\,\,\,,\label{movins2}
\end{eqnarray}
where the effective energy momentum tensor $T_{\mu\nu}^{eff}$ is defined by
\begin{eqnarray}
T_{\mu\nu}^{eff}=\frac{1}{f_{1R}(R)}\Big\{T_{\mu\nu}+f_{2T}(T)T_{\mu\nu}+\frac{1}{2}\left[ 2f_{2T}(T)p+f_2(T)+f_1(R)-Rf_{1R}(R)\right] g_{\mu\nu}\nonumber\\-\left( g_{\mu\nu}\Box-\nabla_\mu\nabla_\nu\right) f_{1R}(R)\Big\}\,\,\,.\label{nouvelle3}
\end{eqnarray} 
Since the vector field $k^{\mu}$ is lightlike, one gets $k^{\mu}k_{\mu}=0$. Then, by multiplying Eq.~(\ref{movins2}) by the bi-vector $k^{\mu}k^{\nu}$, the null energy condition comes to be equivalent to $T_{\mu\nu}^{eff}k^{\mu}k^{\nu}\geq 0$. Remark also that Eq.~(\ref{movins2}) is equivalent to 
\begin{eqnarray}
R_{\mu\nu}=T^{eff}_{\mu\nu}-\frac{1}{2}T^{eff}g_{\mu\nu}\,\,\,.
\end{eqnarray}
Then, the strong energy condition reads $\left(T^{eff}_{\mu\nu}-\frac{1}{2}T^{eff}g_{\mu\nu}\right)V^{\mu}V^{\nu}\geq 0$. By assuming that the total content of the universe behaves as perfect fluid, Eq.~(\ref{vincent5}) also holds, by just substituting $\rho$ and $p$, by $\rho_{eff}$ and $p_{eff}$ respectively (the effective energy density and effective pressure). Thus, the null energy condition imposes
\begin{eqnarray}
T_{eff}^{\mu\nu}k^{\mu}k^{\nu}=\left(\rho_{eff}+p_{eff}\right)u_{\mu}u_{\nu}
k^{\mu}k^{\nu}-p_{eff}g_{\mu\nu}k^{\mu}k^{\nu}\geq 0\,\,\,.
\end{eqnarray}
Since, $k^{\mu}$  is a lightlike  vector, one has $g_{\mu\nu}k^{\mu}k^{\nu}=k^{\mu}k_{\mu}=0$. Also, we  always get $u_{\mu}u_{\nu}k^{\mu}k^{\nu}=(u_{\mu}k^{\mu})^2>0$. Thus, the null energy condition for the effective perfect fluid reduces to 
\begin{eqnarray}
\rho_{eff}+p_{eff}\geq 0\,\,\,.\label{nullcondition}
\end{eqnarray} 
For the strong energy condition, one has
\begin{eqnarray}
\left(T_{\mu\nu}-\frac{1}{2}Tg_{\mu\nu}\right)V^{\mu}V^{\nu}&=&\left[\left(\rho_{eff}+ p_{eff}\right)u_{\mu}u_{\nu}-p_{eff}g_{\mu\nu}-\frac{1}{2}\left(\rho_{eff}- 3p_{eff}\right)g_{\mu\nu}\right]V^{\mu}V^{\mu}\geq 0\,\,\,,\label{grande}\\
&=&\left(\rho_{eff}+3p_{eff}\right)V^{\mu}V_{\mu}
\end{eqnarray}
where we made use of $T^{eff}=\rho_{eff}-3p_{eff}$ and $  u_{\mu}u_{\nu}V^{\mu}V^{\nu}=V_{\mu}V^{\mu}$ for obtaining the second equality.\footnote{ Since $u_{\mu}$ is an unitary timelike vector (four-velocity) and the metric component $g_{00}=g^{00}=1$ (see Eq.~(\ref{vincent9})), one gets $u_{\mu}u_{\nu}V^{\mu}V^{\nu}=V_{\mu}V^{\mu}$.} Remark that the condition (\ref{grande}) is equivalent to
\begin{eqnarray}
\left(\rho_{eff}+ p_{eff}\right)V^{\mu}V_{\mu}\geq \frac{1}{2}\left(\rho_{eff}- p_{eff}\right)V^{\mu}V_{\mu}\,\,\,,
\end{eqnarray} 
which imposes $\left(\rho_{eff}+ p_{eff}\right)V^{\mu}V_{\mu}\geq 0$. Since the vector $V^{\mu}$ is timelike, one has $V^{\mu}V_{\mu}\geq 0$, and the strong energy condition reduces to 
\begin{eqnarray}
\rho_{eff}+p_{eff}\geq 0\,\,\,,\quad\quad \rho_{eff}+3p_{eff}\geq 0\,\,\,\,.
\end{eqnarray}
Note that the condition $\left(T^{eff}_{\mu\nu}-\frac{1}{2}T^{eff}g_{\mu\nu}\right)V^{\mu}V^{\nu}\geq 0$ is equivalent to $T^{eff}_{\mu\nu}V^{\mu}V^{\nu}\geq \frac{1}{2}T^{eff}V^{\mu}V_{\nu}$. This requires to have  the necessary condition $T^{eff}_{\mu\nu}V^{\mu}V^{\nu}\geq 0$.
The weak energy condition requires to this condition  holds for any non-spacelike vector \cite{hawking,sayan}. This means that the condition holds for both lightlike and timelike vectors.  The use of the lightlike vector is equivalent to the null energy condition as previously demonstrated. Now, for a timelike vector, one gets
\begin{eqnarray}
T^{eff}_{\mu\nu}V^{\mu}V^{\nu}&=&\left[\left(\rho_{eff}+p_{eff}\right)u_{\mu}u_{\nu}-p_{eff}g_{\mu\nu}\right]V^{\mu}V^{\nu}\geq 0\,\,\,,\nonumber\\
&=&\rho_{eff}V^{\mu}V_{\mu}\geq 0\,\,\,.
\end{eqnarray}
Since, $V^{\mu}$ is a timelike vector, one always has $V^{\mu}V_{\mu}\geq 0$. Thus, the weak energy condition for the effective perfect fluid reads
\begin{eqnarray}
\rho_{eff}\geq 0\,\,\,,\quad\quad\quad \left(\rho_{eff}+p_{eff}\right)\geq 0\,\,\,.
\end{eqnarray}
For the dominant energy condition, the required condition is that, besides  the weak energy condition, for any timelike vector $V^{\mu}$, $T_{\mu\nu}V^{\mu}$ is a non-spacelike vector. In other word, this means that no signal can propagate faster than light. It follows that $p_{eff}\leq \rho_{eff}$, since, for any know matter, pressures are small  when the energy density is small. Thus, the dominant energy condition results in
\begin{eqnarray}
\rho_{eff}\geq 0\,\,,\quad\quad \rho_{eff}+p_{eff}\geq 0\,\,,\quad\quad\quad \rho_{eff}-p_{eff}\geq 0\,\,\,.
\end{eqnarray}

Therefore, the energy conditions, as known in GR, can also be applied in this modified theory of gravity by substituting the ordinary energy density $\rho$ and pressure $p$ in GR by the effective ones, $\rho_{eff}$ and $p_{eff}$.\par
In what follows, we will consider models of type $f(R, T)=R+2f(T)$, i.e., the usual Einstein-Hilbert term plus trace depending term  $2f(T)$. This amounts to consider $f_1(R)=R$ and $f_2(T)=2f(T)$. The factor $2$ is used just for letting the field equations more easier to be treated. We will also assume that the ordinary content of the universe is pressureless and satisfies the energy conditions (just $\rho\geq 0$).

\section{Testing some $f(R, T)= R+2f(T)$ models from energy conditions}
In this section we will present the conditions required on $\rho$ and the algebraic function $f(T)$ for realizing each type of energy conditions. For this end, we first need to establish the respective expression of the effective energy density $\rho_{eff}$ and effective pressure $p_{eff}$. According to the assumptions made at the end of the previous section, Eq.~(\ref{vincent7}) becomes
\begin{eqnarray}\label{vincent8}
R_{\mu\nu}-\frac{1}{2}Rg_{\mu\nu}=T_{\mu\nu}+ 2f_{T}(T)T_{\mu\nu}+f(T)g_{\mu\nu}\,\,.
\end{eqnarray}
Considering the flat FRW space-time described by the metric
\begin{eqnarray}\label{vincent9}
ds^2=dt^2-a^2(t)d{\bf x}^2\,\,,
\end{eqnarray}
where $a(t)$ is the scale factor.  The $00$ and $ii$ components of (\ref{vincent8}) can be written as 
\begin{eqnarray}
3H^2&=& \rho_{eff}\label{vincent10} \,\,,\\
-2\dot{H}-3H^2&=& p_{eff}\label{vincent11}\,\,,
\end{eqnarray}
where the effective energy density and pressure are defined as
\begin{eqnarray}
\rho_{eff}&=& \rho +2\rho f_{T}(T)+f(T)\,\,,\label{vincent12}\\
p_{eff}&=& -f(T)\,\,.\label{vincent13}
\end{eqnarray} 
By using the above expressions of the effective energy density and pressure, we get the null energy condition (NEC), the weak energy condition (WEC), the strong energy condition (SEC) and the dominant energy condition (DEC) by\par

\begin{eqnarray}
\mbox{NEC:}\quad\quad\quad\quad\quad\quad\rho_{eff}+p_{eff}= \rho\left[1+2f_{T}(T)\right]\geq 0 \,\,\,,\label{vincent14}
\end{eqnarray}
\begin{eqnarray}
\mbox{WEC:}\quad\quad\quad\quad\quad\quad \rho_{eff}= \rho +2\rho f_{T}(T)+f(T)\geq 0\,\,,\quad \rho_{eff}+p_{eff}\geq 0\,\,\,,\label{vincent15}
\end{eqnarray}
\begin{eqnarray}
\mbox{SEC:}\quad\quad\quad\quad\rho_{eff}+3p_{eff}=\rho+2\rho f_{T}(T)-2f(T)\geq 0 \,\,,\quad \rho_{eff}+p_{eff}\geq 0\,\,,\label{vincent16}
\end{eqnarray}
\begin{eqnarray}
\mbox{DEC:}\quad\quad\quad\rho_{eff}-p_{eff}=\rho+2\rho f_{T}(T)+2f(T)\geq 0\,\,,\quad \rho_{eff}+p_{eff}\geq 0\,\,,\quad \rho_{eff}\geq 0\,\,.\label{vincent17}
\end{eqnarray}
We propose to test two models of $f(T)$ in the way to make them satisfying the  energy conditions.                                                                                                                                                                                                                                                                                                    For the first model, we consider a function $f(T)$ such that  for large and small values of the trace, it converges. We start by assuming first this function as $f(T)=\frac{a_1T^n+b_1}{a_2T^n+b_2}$, where $a_1$, $a_2$, $b_1$, $b_2$ and $n$ are parameters to be adjusted in order to obtain models that satisfy some or all the energy conditions. Note that this expression is similar to that used in \cite{model1,model2}, where, at the place of the trace $T$, the Gauss-Bonnet invariant is used. A motivation in using this model is that when $n>0$, it prevents  divergence for large and small values of the trace\footnote{This requires to fix $a_2\neq 0$ for large values of trace and $b_2\neq 0$ for small values of the trace.}. Remark that for $T\rightarrow \infty$, $f(T)=a_1/a_2$ is finite, and for $T\rightarrow 0$, $f(T)=b_1/b_2$ which is also finite. However, when $n<0$, the situation inverts, and for large and small values of the trace\footnote{This requires to fix $a_2\neq 0$ for small values of trace and $b_2\neq 0$ for large values of the trace.}, one obtains  $f(T)=b1/b2$ and $f(T)=a1/a2$, respectively, and the divergence is still prevented. But note that all this is a mathematical point of view, which also may be useful at the physical point of view. Let us now explore the cosmological feature of this model. Here we can discuss the occurrence of cosmic acceleration, which may impose some constraints to the parameters, reducing the degree of freedom of the model. Searching for the power-like solutions, one has: $a=a_0t^x$, $H=xt^{-1}$, $\dot{H}=-xt^{-2}$ and $f(T)\rightarrow g(t)=(\bar{a}_1t^{-3xn}+b_1)/(\bar{a}_2t^{-3xn}+b_2)$, with $\bar{a}_i=a_i\rho_0a_0^{-3}$ ($i=1,2$), where we used $T=\rho=\rho a^{-3}$ (because the ordinary content of the universe is dust). With the scale factor being used here, the early universe corresponds to $t\rightarrow 0$ and the late time correspond to $t\rightarrow \infty$, where $x>0$. At early time, the curvature scalar $R\rightarrow \infty$ (where the $f(T)$ contribution may be neglected) and at late-time, the universe may be characterized by the $\Lambda CDM$ model, i.e., our model  must behave like $R+2\Lambda$. For obtaining this feature, for $n<0$, the cosmological constant reads $\Lambda = \bar{a}_1/\bar{a}_2=a_1/a_2$, while for $n>0$, one gets $\Lambda = {b}_1/{b}_2$. By calling $\omega_{eff}$ the effective parameter of the equation of state, one has $\omega_{eff}={p_{eff}}/\rho_{eff}=-f(T)/(\rho+2\rho f_\rho+f(T))$. The requirement  of guaranteeing the acceleration of the universe, without falling into phantom model, is $-1<\omega_{eff}<-1/3$. Observe that at time, ($t\rightarrow \infty$), $\omega_{eff}=-1$ (this holds only when $n>0$). Thus, we see that some values of the input parameters can make the model providing the late time acceleration, i.e., for $n>0$, $a_1/a_2>0$ and $b_1/b_2=\Lambda$. It is important to note that when the acceleration is guaranteed, at least the strong energy condition is violated. But here, we recall that the aim of this work is to find the range of  the parameters for which the energy conditions are satisfied. We will develop this in the subsection ${\bf 4.1}$. \par
For the second model, we assume $f(T)=a_3\ln^{q}{\left(b_3T^m\right)}$  where $a_3$, $m$ and $q$ are also adjustable parameters and $b_3$ is assumed to be positive and non null. This form is chosen due to its interesting aspect, because it is a type of the $f(T)$ model for which, the future finite-time singularity of type-I (the Big Rip) can be cured in $f(R, T)=R+2f(T)$ gravity \cite{juliano}. Despite the aim of this work not being the study of the avoidance of singularities, we can try to show a little more how this type of model can prevent the Big Rip. Remember that the Big Rip is the type of the future finite-time singularity for which, the scale factor, the energy density and the pressure diverge. This also implies that at the singularity time, here denoted by $t_s$, the curvature scalar $R$ diverges. Then, in order to show the avoidance of the Big Rip from the contribution of the $f(T)$, let us take the trace of the Eq.~(\ref{vincent8}), i.e.
\begin{eqnarray}
-R=\rho+2\rho f_T(T)+4f(T)\,\,\,.\label{trace2}
\end{eqnarray}
Remark that in \cite{juliano}, the Hubble parameter is written as $H=h(t_s-t)^{-\alpha}$, where $h$ is a positive real parameter and $\alpha>1$; this is the condition for the appearance of the Big Rip. In this case, the scale factor reads $a=\bar{a}exp{\left[h(t_s-t)^{1-\alpha}/(\alpha-1)\right]}$, where $\bar{a}$ is a positive parameter. Hence, as the singularity is approached, the curvature $R$  behaves like $(t_s-t)^{-2\alpha}$. On the other hand, one can determine the quantity $b_3T^m=b_3\rho^m=b_3\bar{\rho}^m\bar{a}^{-3m}\exp{\left[3hm(t_s-t)^{1-\alpha}/(1-\alpha)\right]}$ and the $f(T)=f(\rho)$ behaves like $(t_s-t)^{q(1-\beta)}$. From these points, with $m<0$,  we see that for $q>2\alpha/(\alpha-1)$, $q>0$ and $q(1-\alpha)<0$ because of $\alpha>1$. Moreover, we can see that $q(1-\alpha)<-2\alpha$. Hence, as the singularity is approached, the $T$ contribution diverges more than the curvature scalar, and the Big Rip may be avoided, without the need of quantum effects. This proves that the model may allow itself the avoidance of the Big Rip. Here, we propose to determine the range for the parameters for which the energy conditions are  satisfied, and guaranteeing the auto avoidance of the Big Rip. Note also that in \cite{juliano} it is demonstrated that this model can provide the late time accelerated expansion of the universe.
\par

\subsection{Studying the case $f(T)=\frac{a_1T^n+b_1}{a_2T^n+b_2}$}\label{function1}
Our task here is to put out the constraints on the input parameters in order to get a  $R+2f(T)$ type model that satisfies the energy conditions. According to the sign of the parameter $n$, and assuming that $a_2$ and $b_2$ cannot be identically null, the model can be cast into two different forms. In fact, for the late time stage of the universe,  by dividing the parameters of the model by $a_2$ ($n>0$) and $b_2$ ($n<0$), one gets respectively the models $f(\rho)=(\Lambda \rho^n+B_1)/(\rho^n+B_2)$ and $f(\rho)=(A_1\rho^n+\Lambda)/(A_2\rho^n+1)$, where the cosmological constant is characterized by $a_1/a_2$ (for $n>0$) and $b_1/b_2$ (for $n<0$), and $A_1=a_1/b_2$, $A_2=a_2/b_2$, $B_1=b_1/a_2$ and $B_2=b_2/a_2$. In this case, the model which initially was four parameters dependent, under the cosmological constraints, becomes three parameters dependent, $\Lambda$, $B_1$ and $B_2$ for $n>0$, and $\Lambda$, $A_1$ and $A_2$ for $n<0$.  Since the cosmological constant is known \cite{barrow}\footnote{The cosmological constant is positive and is  $\Lambda \sim 1.7\times 10^{-121}$ Planck units.}, the model turns into two parameters dependent.

\par
The first derivative of $f(T)$ with respect to $T$ (or the derivative of $f(\rho)$ with respect to $\rho$) reads
\begin{eqnarray}\label{vincent18}
f_\rho(\rho)=\left\{\begin{array}{ll}
\frac{n(\Lambda B_2-B_1)\rho^{n-1}}{(\rho^n+B_2)^2},\quad \mbox{for}\,\,\, n>0 \\
\frac{n(A_1-\Lambda A_2)\rho^{n-1}}{(A_2\rho^n+1)^2}, \quad \mbox{for}\,\,\, n<0\,\,\,.\end{array}\right.
\end{eqnarray}

$\bullet$ The NEC\par

Since we have assumed that the ordinary content of the universe satisfies all the energy conditions,  the condition (\ref{vincent14})  reduces to $1+2f_{T}(T)\geq 0$,   (or  $1/2+f_{\rho}(\rho)\geq 0$). One can calculate $1/2+f_{\rho}(\rho)$ as

\begin{eqnarray}\label{vincent19}
f_{\rho}(\rho)+\frac{1}{2}= \left\{\begin{array}{ll}
\frac{2n\rho^{n-1}\left(\Lambda B_2-B_1\right)+\left(\rho^n+B_2\right)^2}{2\left(\rho^n+B_2\right)^2}\,\,\,,\quad \mbox{for}\,\,\, n>0\\
\frac{2n\rho^{n-1}\left(A_1-A_2\Lambda\right)+\left(A_2\rho^n+1\right)^2}{2\left(A_2\rho^n+1\right)^2}\,\,\,, \quad \mbox{for}\,\,\, n<0\,\,\,\,,
\end{array}\right.
\end{eqnarray}
whose the sign can just be characterised by that of the numerator, since the denominator is always positive. If we take the numerator as a function of the ordinary energy density $\rho$ and the input parameters, we just need to analyse the sign of this latter. The evident conditions\footnote{Here, we call ``evident conditions" the conditions based just on the signs of the parameter without the need of knowing their absolute values.}  for which the numerator is positive are presented as follows:\par
$\ast$   $B_1>0$, \quad $B_2>0$,  \quad $ B_1/B_2<\Lambda$\quad  for \quad $n>0$\,\,\,,
\par 
$\ast$  $A_1>0$, \quad $A_2> 0$,  \quad $ B_1/B_2>\Lambda$ \quad for \quad  $n<0$\,\,\,,
\par
$\ast$  $B_1<0$, \quad $B_2>0$,  \quad for \quad $n>0$\,\,\,,

\par

$\ast$ $A_1<0$, \quad $A_2> 0$, \quad for \quad $n<0$\,\,\,.
\par

Indeed, the above conditions lead to the positivity  $2n\rho^{n-1}(\Lambda B_2-B_1)>0$ for $n>0$ and $2n\rho^{n-1}(A_1-A_2\Lambda)>0$ for $n<0$. Observe that there are still situations in which the above quantities are negative but the numerators in (\ref{vincent19}) continuing positive, i.e.,
\par
$\ast$ $ A_1>A_2\Lambda$\quad and $|2n\rho^{n-1}\left(A_1-A_2\Lambda\right)|<|\left(A_2\rho^n+1\right)^2|$ for $n<0$\,\,,\par
$\ast$ $\Lambda B_2<B_1$ and $|2n\rho^{n-1}\left(\Lambda B_2-B_1\right)|<|\left(\rho^n+B_2\right)^2|$\quad for \quad $n>0$\,\,.\par
In these cases, one can plot the function in terms of two of the parameters, fixing the other. Despite knowing the sign of the considered parameters with what respect the function may be plotted, the important here is their rank, i.e. the interval to which they must belong in order to produce the positivity of the function. Some examples are presented in Fig. 1.\par

$\bullet$ The WEC\par
This condition is realized when the NEC is, plus the condition $\rho_{eff}\geq 0$. Note that the complete expression and condition of the NEC read
\begin{eqnarray}\label{vincent20}
2n\left(\Lambda B_2-B_1\right)\rho^n+\rho^{n+1}+2B_2\rho^{n+1}+B_2^2\rho \geq 0\,\,,\quad \mbox{for} \quad n>0\,\,,\\
2n\left(A_1-A_2\Lambda\right)\rho^n+A_2^{2}\rho^{n+1}+2A_2\rho^{n+1}+\rho \geq 0\,\,,\quad \mbox{for} \quad n<0\,.\label{vincent201}
\end{eqnarray}
These expressions are obtained by multiplying the numerators in (\ref{vincent19}) by $\rho$. We didn't need to use this complete expression for determining the conditions on the input parameters in the case of the NEC,  since the ordinary energy density is assumed as positive quantity. Besides to (\ref{vincent20})-(\ref{vincent201}), the second condition for satisfying the WEC is 
\begin{eqnarray}\label{vincent21}
\rho_{eff}=\rho+2\rho f_{\rho}+f(\rho)\geq 0\quad,
\end{eqnarray}
having in  mind that the ordinary content is assumed as pressureless. By using $f(\rho)$ according to the functions in (\ref{vincent18}), (\ref{vincent21}) becomes
\begin{eqnarray}\label{vincent22}
\Lambda \rho^{2n}+\rho^{2n+1}+2B_2\rho^{n+1}+\left(2n\Lambda B_2-2nB_1
+\Lambda B_2+B_1\right)\rho^n+B_2^2 \rho+B_1B_2 \geq 0 \,\,, \mbox{for}\;\; n>0,\\
A_1A_2\rho^{2n}+A_2^2\rho^{2n+1}+2A_2\rho^{n+1}+\left(2nA_1-2nA_2\Lambda
+A_1+\Lambda A_2\right)\rho^n+ \rho+\Lambda \geq 0 \,\,,\mbox{for}\;\;n<0.\label{vincent221}
\end{eqnarray}
Note here that we just use the numerator of the fractions whose the denominators are always positives.
By combining (\ref{vincent20}) with (\ref{vincent22})-(\ref{vincent221}), one gets for the WEC
\begin{eqnarray}\label{vincent23}
\Lambda\rho^{2n}+2\rho^{2n+1}+4B_2 \rho^{n+1}+2B^2_2\rho+\left[(1+4n)\Lambda B_2+(1-4n)B_1\right]\rho^n+B_1B_2 \geq 0,\;\;\mbox{for}\;\; n>0\,,\\
A_1A_2\rho^{2n}+2\rho^{2n+1}+4A_2 \rho^{n+1}+2\rho+\left[(1+4n)A_1+(1-4n)A_2\Lambda\right]\rho^n+\Lambda  \geq 0,\;\;\mbox{for}\;\; n<0\,\,\,. \label{vincent231}
\end{eqnarray}
We address here the evident conditions for which the WEC is satisfied as follows:
\par
$\ast$  $B_1>0$, \quad $B_2>0$\quad for \quad $0<n<1/4$\,\,\,,

$\ast$ $A_1>0$,\quad $A_2>0$, \quad for \quad $-1/4<n<0$\,\,\,.

It is obvious that these conditions are not unique. For $n>0$ ($n<0$), the necessity of plotting the function $\Lambda\rho^{2n}+2\rho^{2n+1}+4B_2 \rho^{n+1}+2B^2_2\rho+\left[(1+4n)\Lambda B_2+(1-4n)B_1\right]\rho^n+B_1B_2$,\,\,\, ($A_1A_2\rho^{2n}+2A^2_2\rho^{2n+1}+4A_2 \rho^{n+1}+2\rho+\left[(1+4n)A_1+(1-4n)A_2\Lambda\right]\rho^n+\Lambda $) varying two of the input parameters and fixing the other, appears when at least one of the following expressions is negative, ($B_2$, $B_1B_2$ or $(1+4n)\Lambda B_2+(1-4n)B_1$),\;\;  ($A_1A_2$,  $A_2$,  or $(1+4n)A_1+(1-4n)A_2\Lambda$). We present some examples of theses cases in Fig. 2. \par

$\bullet$ The SEC\par
The strong energy condition is realized by combining the NEC with $\rho_{eff}+3p_{eff}\geq 0$. This latter reads,
\begin{eqnarray}\label{vincent24}
\rho_{eff}+3p_{eff}=\rho+2\rho f_{\rho}-2f(\rho)\geq 0\quad.
\end{eqnarray}
Making use of the expressions in (\ref{}), one obtains a fraction whose the denominator is always positive and the numerator reads
\begin{eqnarray}\label{vincent25}
\rho^{2n+1}+2B_2\rho^{n+1}+B_2^2\rho-2\left[(1-n)\Lambda B_2+(1+n)B_1\right]\rho^n-2\Lambda\rho^{2n}-2B_1B_2 \geq 0\,, \mbox{for}\; n>0\,\,,\\
A_2^2\rho^{2n+1}+2A_2\rho^{n+1}+\rho-2\left[(1-n)A_1+(1+n)A_2\Lambda\right]\rho^n-2A_1A_2\rho^{2n}-2\Lambda \geq 0\, \mbox{for}\; n<0\,\,.
\label{vincent251}
\end{eqnarray}
Now, combining (\ref{vincent25})-(\ref{vincent251}) with the NEC, on gets the following conditions for the SEC
\begin{eqnarray}\label{vincent26}
2\rho^{2n+1}-2\Lambda\rho^{2n}+4B_2\rho^{n+1}+2B^2_2\rho-
2\left[(1-2n)\Lambda B_2+(1+2n)B_1\right]\rho^n-2B_1B_2 \geq 0\,,\; \mbox{for}\; n>0\,,\\
2A_2^2\rho^{2n+1}-2A_1A_2\rho^{2n}+4A_2\rho^{n+1}+2\rho-
2\left[(1-2n)A_1+(1+2n)A_2\Lambda \right]\rho^n-2\Lambda \geq 0\,,\; \mbox{for}\; n<0\,.
\end{eqnarray}

In this case, there is any obvious condition for satisfying the SEC. However, values can be found, by plotting  $2\rho^{2n+1}-2\Lambda\rho^{2n}+4B_2\rho^{n+1}+2B^2_2\rho-
2\left[(1-2n)\Lambda B_2+(1+2n)B_1\right]\rho^n-2B_1B_2 $ \,and\, $2A_2^2\rho^{2n+1}-2A_1A_2\rho^{2n}+4A_2\rho^{n+1}+2\rho-
2\left[(1-2n)A_1+(1+2n)A_2\Lambda \right]\rho^n-2\Lambda$ in terms of two of the parameters. Some examples for illustrating some of these cases are presented in Fig. 3.\par

$\bullet$ The DEC\par
The dominant energy condition is characterized by the WEC combined  with $\rho_{eff}-p_{eff} \geq 0$. Following the same steps as in the previous cases, one easily obtains the DEC as 
\begin{eqnarray}\label{vincent27}
\rho^{2n+1}+\Lambda\rho^{2n}+2B_2\rho^{n+1}+B_2^2\rho+B_1B_2
+\left[(2n+1)\Lambda B_2+(1-2n)B_1\right]\rho^n \geq 0\, \mbox{for}\; n>0\,\,,\\
A^2_2\rho^{2n+1}+A_1A_2\rho^{2n}+2A_2\rho^{n+1}+\rho+\Lambda 
+\left[(2n+1)A_1+(1-2n)\Lambda A_2\right]\rho^n \geq 0\, \mbox{for}\; n<0\,\,.\label{vincent271}
\end{eqnarray}
The evident conditions read \par
$\ast$  $B_1>0$, \quad $B_2>0$\quad and \quad $n>1/2$\,\,.\par

Evidently, other conditions may lead to the accomplishment of the DEC, but, only plotting the functions in (\ref{vincent27})-(\ref{vincent271}). This can occur when at least one of the following expressions is negative, \quad $B_2$, \quad $B_1B_2$, \quad  $(2n+1)\Lambda B_2+(1-2n)B_1$ for (\ref{vincent27}). We recall that the negative parameters must be suitably chosen without loosing the positivity of the expressions (\ref{vincent27})-(\ref{vincent271}). We present some of theses cases in Fig. 4.

\subsection{Studying the case $f(T)=a_3\ln^{q}{\left(b_3T^m\right)}$} \label{function2}
Here we will work with the fundamental conditions for which the model allows the avoidance of the Big Rip.  So, we  propose to check if the range of parameters for which the singularity may be cured can also make the model satisfying the energy conditions. Here, the first derivative of $f(\rho)$ also plays an important role. Deriving $f(\rho)$ with respect to the energy density $\rho$, one gets
\begin{eqnarray}\label{vincent28}
f_{\rho}(\rho)=\frac{qma_3}{\rho}\ln^{q-1}{\left(b_3\rho^m\right)}\quad.
\end{eqnarray}
We believe that each step of constructing the four energy conditions is now clear and we simply present the results and comments as follows:\par

$\bullet$ NEC
\begin{eqnarray}\label{vincent29}
\rho+2q m a_3+\rho\ln^{q-1}{\left(b_3\rho^m\right)} \geq 0\,\,\,\,.
\end{eqnarray}
The evident conditions for obtaining this are \par
$\ast$ $q>0$, \quad $m<0$, \quad $a_3<0$, with $\rho^m>1/b_3$ \,\,\,.\par
 It is important to note that this list is not exhaustive, since in other conditions different from the above ones, the NEC could still be realized. This is a set of the situations where 
for example $2qma_3<0$, but its absolute value is less than the absolute value of $\rho+\rho\ln^{q-1}{(b_3\rho^m)}$, i.e., $|2qma_3|< |\rho+\rho\ln^{q-1}{(b_3\rho^m)}|$. This situation requires knowing some intervals to which the parameters must belong. We present this feature by plotting the function corresponding to the expression (\ref{vincent29}) in terms of some of the input parameters fixing the other. See the graph at left side in Fig. 5.

$\bullet$ WEC
\begin{eqnarray}\label{vincent30}
2\rho+2 q m a_3+ \left(\rho+2qma_3 \right)\ln^{q-1}{\left(b_3\rho^m\right)}+a_3\ln^{q}{\left(b_3\rho^m\right)} \geq 0  \,\,\,\,.
\end{eqnarray}
In this case by  plotting the function (\ref{vincent30}), the WEC can be realized graphically.  This is the set of situations where one of the term in the sum (\ref{vincent30}) is negative, but  it absolute value is less that the absolute value of the sum of the other. An example for this is when $2qma_3<0$, but $|2qma_3|<2\rho+ |\left(\rho+2qma_3 \right)\ln^{q-1}{\left(b_3\rho^m\right)}+a_3\ln^{q}{\left(b_3\rho^m\right)}|$. See the graph at right side in Fig. 5.\par

$\bullet$ SEC
\begin{eqnarray}\label{vincent31}
2\rho+2qma_3+\left( \rho+2qma_3\right)\ln^{q-1}{\left(b_3\rho^m\right)}-2a_3\ln^{q}{\left(b_3\rho^m\right)} \geq 0\,\,\,.
\end{eqnarray}
In this case, evident constraints on the input parameters in order to realize this energy conditions are presented as follows:\par
$\ast$ $q>0$, \quad $m<0$, \quad $a_3<0$, with $\rho^m>1/b_3$ \,\,\,,\par
As presented in the previous cases, other conditions may also realize this energy conditions. This can be observed by plotting the function in (\ref{vincent31}) in terms of some input parameters, fixing the other. See the graph at left side in Fig. 6.\par

$\bullet$ DEC
\begin{eqnarray}\label{vincent32}
3\rho+2qma_3+\left(\rho+4qma_3\right)\ln^{q-1}{\left(b_3\rho^m\right)}+3a_3\ln^{q}{\left(b_3\rho^m\right)} \geq 0\,\,\,.
\end{eqnarray}
 Here, constraints may also lead to the DEC, but this is clear  by  plotting the function (\ref{vincent32}), as in the previous cases. We present an illustrative example at right side in Fig. 6.\par

\par
We mention that for all the graphs, the parameters are  normalised  to $10^{-121}$ Planck units. Remark that the current value of the cosmological constant is about $1.7\times 10^{-121}$ and the energy density of the usual matter is about $\rho=0.1\times 10^{-121}$ \cite{barrow}. Then, with the normalization, we get $\Lambda=1.7$ and $\rho=0.1$ for the cosmological constant and the energy density of the usual matter respectively, which are the values used for plotting the graph in the figures. Moreover, we observe that all the conditions which lead to the accomplishment of the NEC also satisfy the WEC, the SEC and the DEC. This is an obvious deduction from (\ref{vincent14})-(\ref{vincent17}). In the same way, it is easy to observe that all the conditions for  which the WEC is satisfied, also lead to the accomplishment of the DEC. From the same expressions (\ref{vincent15})-(\ref{vincent16}), we see that the sufficient condition for the requirement of satisfying the WEC leads to the accomplishment of the SEC is $f(T)< 0$. Thus, among the conditions for which the WEC is satisfied, those for which the function is always negative, also lead to the accomplishment of the SEC. However, about the SEC with respect to the DEC, the situation is quite different. The requirement for, the conditions which satisfy the SEC, leading to the accomplishment of the DEC, is always having $f(T)>0$. This means that, among the conditions for which the SEC is satisfied, those for which the function $f(T)$ is always positive, guarantee the accomplishment of the DEC.
\section{Perturbations and stabilities in $R+2f(T)$ gravity}
In this section we propose to study the perturbations around the models used in this work. We can start  establishing  the perturbed equations for the case $R+2f(T)$, but the two models will be studied as specific cases.\par
For this purpose, let us assume a general solution for the cosmological background of FRW metric, which is given by a Hubble parameter $H = H_b(t)$ that satisfies the background equation (\ref{vincent12}) using (\ref{vincent10}), for $R+2f(T)$ gravity. The evolution of the matter energy density can be expressed in terms of this particular solution by solving the continuity equation around $H_b(t)$,
\begin{eqnarray}
\dot{\rho}_b+3H_b(t)\rho_b=0\,\,\,,
\end{eqnarray} 
yielding 
\begin{eqnarray}
\rho_b(t)=\rho_0e^{-3\int H_b(t)dt}\,\,\,\,.
\end{eqnarray}
We recall that we are considering that the ordinary content of the universe is pressureless. Since we are interesting in studying the perturbations around the solutions $H=H_b(t)$, we will consider small deviations from the Hubble parameter and the energy density, i.e., we can write the Hubble parameter and the ordinary energy density as \cite{diego}
\begin{eqnarray}
H(t)=H_b(t)\left(1+\delta(t)\right)\,\,\,,\quad \rho(t)=\rho_b(t)\left(1+\delta_m(t)\right)\,\,\,. \label{perturbationdef}
\end{eqnarray}
In order to study the behaviour of these perturbations in the linear regime, we expand the function $f(T)$ in powers of $T_b$ (or $\rho_b$) evaluated at the solution $H=H_b(t)$, as
\begin{eqnarray}
f(T)=f(\rho)=f^b+f_\rho^b(\rho-\rho_b)+\mathcal{O}^2\,\,\,\,,\label{functionfT}
\end{eqnarray}
where the superscript $b$ refers to the background values of $f(T)$ and its derivatives evaluated at $T=T_b$ (or $\rho=\rho_b$). Here, the $\mathcal{O}$ term includes all the terms proportional to the square or higher powers of $T$ (or $\rho$). Then, only the linear terms of the induced perturbations will be considered. Hence, by making use of the expression (\ref{perturbationdef}) in the Eqs.~(\ref{vincent12}) and (\ref{vincent10}), one gets the equation for the perturbation $\delta(t)$ in the linear approximation,
\begin{eqnarray}
\left(\rho_b+3\rho_bf^b_{\rho}+2\rho^2_bf^b_{\rho\rho}\right)\delta_m(t)=6H_b^2\delta(t)\,\,\,.\label{firstequation}
\end{eqnarray}
On the other hand, there is a second perturbed equation from the matter continuity equation,
\begin{eqnarray}
\dot{\delta}_m+3H_b(t)\delta(t)=0\,\,\,. \label{secondequation}
\end{eqnarray} 
By combining Eqs.~(\ref{firstequation}) and (\ref{secondequation}) one gets the following equation for the matter perturbation
\begin{eqnarray}
2H_b\dot{\delta}_m+\left(\rho_b+3\rho_bf^b_{\rho}+2\rho^2_b f^b_{\rho\rho}\right)\delta_{m}=0\,\,\,,
\end{eqnarray}
from which we obtain
\begin{eqnarray}
\delta_m(t)=C_1\exp{\left\{-\frac{1}{2}\int C_b dt\right\}}\,\,, \quad C_b=\frac{\rho_b}{H_b}\left(1+3f^b_{\rho}+2\rho_bf^b_{\rho\rho}\right)\,\,\,,\label{firstintegral}
\end{eqnarray}
where $C_1$ is an integration constant. By using the relation (\ref{secondequation}), the perturbation $\delta$ reads
\begin{eqnarray}
\delta(t)=-\frac{C_1C_b}{6H_b}\exp{\left\{-\frac{1}{2}\int C_b dt\right\}}\,\,.
\end{eqnarray} 
Let us now consider two cosmological solutions and analyse their stability by the use of the models treated in this work: de Sitter solutions and power-law solutions.
\subsection{Stability of de Sitter solutions}
In de Sitter solutions, the Hubble parameter is constant and one has 
\begin{eqnarray}
H_b(t)=H_0\,\,,\rightarrow a(t)=a_0e^{H_0t}\,\,\,,
\end{eqnarray}
where $H_0$ is constant. \par
With this scale factor, the energy density of the background becomes $\rho_b=\rho_0e^{-3H_0t}$, with which one has  $d\rho_b=-3H_0\rho_bdt$. By using this, one can cast the integral in (\ref{firstintegral}) into 
\begin{eqnarray}
\int C_bdt=-\frac{1}{3H^2_0}\int \left(1+3f^b_\rho+2\rho_bf^b_{\rho\rho}\right)d\rho_b\,\,\,.\label{secondintegral}
\end{eqnarray}
$\bullet$ {\bf Treating the model $f(T)=\frac{\Lambda T^n+B_1}{T^n+B_2}$}.\par 
This case corresponds to $n>0$, and the integral (\ref{secondintegral}) can be expressed as 
\begin{eqnarray}
\int C_bdt=-\frac{1}{3H^2_0}\left[\rho_b-\frac{\Lambda B_2-B_1}{B_2+\rho^n_b}+\frac{2n(\Lambda B_2-B_1)\rho_b^n}{(B_2+\rho_b^n)^2}\right]\,\,,\quad \rho_b^n=\rho_0^ne^{-3nH_0t}\,\,\,,\label{thirdintegral}
\end{eqnarray}
and $C_b$ is written as
\begin{eqnarray}
C_b=\frac{1}{H_0}\left[\rho_b+\frac{n(2n+1)(\Lambda B_2-B_1)\rho_b^n}{(\rho_b^n+B_2)^2}-\frac{4n^2(\Lambda B_2-B_1)\rho_b^{2n}}{(\rho_b^n+B_2)^3}\right]\,\,.\label{finalCb1}
\end{eqnarray}
We see  from (\ref{thirdintegral}) and (\ref{finalCb1}) that for $n>0$, and as the time evolves, the stability of de Sitter solutions requires $B_2\neq 0$. In other word, for the initial model $f(T)=\frac{a_1T^n+b_1}{a_2T^n+b_2}$, de Sitter solutions are stables if and only if $a_2\neq 0$ and $b_2\neq 0$.\par
$\bullet$ {\bf Treating the model $f(T)=\frac{A_1 T^n+\Lambda}{A_2T^n+1}$}.\par 
This case corresponds to $n<0$, and the integral (\ref{secondintegral}), multiplied by $-1/2$, can be expressed as 
\begin{eqnarray}
-\frac{1}{2}\int C_bdt=\frac{1}{6H^2_0}\left[\rho_b-\frac{A_1-\Lambda A_2}{A_2(1+A_2\rho^n_b)}+\frac{2n(A_1-\Lambda A_2)\rho_b^n}{(1+A_2\rho_b^n)^2}\right]\,\,,\quad \rho_b^n=\rho_0^ne^{-3nH_0t}\,\,\,,\label{thirdintegral2}
\end{eqnarray}
and $C_b$ is written as
\begin{eqnarray}
C_b=\frac{1}{H_0}\left[\rho_b+\frac{n(2n+1)(A_1-\Lambda A_2)\rho_b^n}{(A_2\rho_b^n+1)^2}-\frac{4n^2A_2(A_1-\Lambda A_2)\rho_b^{2n}}{(A_2\rho_b^n+1)^3}\right]\,\,.\label{finalCb12}
\end{eqnarray}
Here, for $n<0$,  as the time evolves, both (\ref{thirdintegral2}) and (\ref{finalCb12}) tend to $+\infty$. Thus the perturbation will grow exponentially, and this particular de Sitter solution becomes unstable. Note that this result does  not depend on any of the parameters $A_1$ or $A_2$.\par
$\bullet$ {\bf Treating the model $f(T)=a_3\ln^q{\left(b_3 T^m\right)}$}.\par
With this model, the integral (\ref{secondintegral}), multiplied by $-1/2$, can be performed and one gets 
\begin{eqnarray}
-\frac{1}{2}\int C_bdt=\frac{1}{6H_0}\left[\rho_b+a_3\ln^q{\left(b_3\rho^m\right)}+
2mqa_3\ln^{q-1}{\left(b_3\rho^m\right)}\right]\,\,\,,\label{model22}
\end{eqnarray}
with the corresponding expression of $C_b$ being
\begin{eqnarray}
C_b=\frac{1}{H_0}\left[\rho_b+qma_3\ln^{q-1}{\left(b_3\rho^m\right)}+2mq(q-1)a_3\ln^{q-2}{\left(b_3\rho^m\right)}\right]\,\,.\label{model23}
\end{eqnarray}
Let us recall that this model ($f(T)=a_3\ln^q{\left(b_3 T^m\right)}$), leads to the avoidance of the Big Rip for $q>2\alpha/(\alpha-1)$ and $m<0$, where $\alpha>1$, as we have previously shown. These conditions also allow the model to satisfy the energy conditions. Now, let us check what happens about the stability with these conditions. First, note that the relation $q>2\alpha/(\alpha-1)$ can be cast into $q>2+2/(\alpha-1)$, showing that $q>2$ because of $\alpha>1$. By choosing $a_3<0$, we see that, within the conditions $q>2$ and $m<0$, the expressions (\ref{model22}) and (\ref{model23}) tend to $-\infty$ as the time evolves, and this ensures the decay of the perturbation, leading to the stability of de Sitter solutions with this model.
Thus,  regarding to the stability of de Sitter solutions, the energy conditions and the late time acceleration, provided with the conditions $a_3<0$, $b_3>0$, $q>2$ and $m<0$, we can conclude that the model may be cosmologically acceptable. 

\subsection{Stability of power-law solutions}
As we are dealing with dust as ordinary content of the universe, we will be interested to the scale factor 
\begin{eqnarray}
a(t)a_0 t^{2/3}\,\,\,\rightarrow \,\,\,H_b(t)=\frac{2a_0}{3t}\,\,\,,\quad \rho_b\propto t^{-2}\,\,\,.
\end{eqnarray}
$\bullet$ {\bf Treating the model $f(T)=\frac{\Lambda T^n+B_1}{T^n+B_2}$}.\par

In this case, $n>0$, and  one can perform the integral
\begin{eqnarray}
-\frac{1}{2}\int C_bdt=-\frac{3}{4a_0}\Big[\ln{(t)}-\frac{(3n+4)(\Lambda B_2-B_1)t^2}{B_2(1+B_2t^{2n})}+\frac{(\Lambda B_2-B_1)t^2}{2B_2n(1+B_2t^{2n})^2}+\nonumber\\
+\frac{(4n+7)(\Lambda B_2-B_1)t^2}{2B_2}\,_2F_1\left(\frac{1}{n},1,1+\frac{1}{n};-B_2t^{2n}\right)\Big]\,\,\,,
\end{eqnarray}
with, 
\begin{eqnarray}
C_b=\frac{3}{2a_0}\left[\frac{1}{t}+\frac{2n(n+1)(\Lambda B_2-B_1)t^{-2n}}{(t^{-2n}+B_2)^2}-\frac{4n^2(\Lambda B_2-B_1)t^{-4n}}{(t^{-2n}+B_2)^3}\right]\,\,,
\end{eqnarray}
where we have set $\rho_0=1$, and $\,_2F_1$ is the hypergeometric function defined by
\begin{eqnarray}
\,_2F_1(\lambda_1,\lambda_2,\lambda_3;z)=\sum^{\infty}_{r=0} \frac{(\lambda_1)_r(\lambda_2)_r}{(\lambda_3)_r}\frac{z^r}{r!}\,\,\,\\
(\lambda)_r=\lambda(\lambda+1)(\lambda+2)...(\lambda+r-1)\,\,\,,\quad (\lambda)_0=1\,\,.
\end{eqnarray}
As the time evolves, conditions are required for guaranteeing the decay of the perturbation. For $B_2<0$, it is necessary to have $\Lambda B_2<B_1$, which means that $B_1$ can be positive, or negative but with $|\Lambda B_2|>|B_1|$. In the case where $B_2>0$ one may observe two sub-cases, i.e., for an even $r$, and an odd  $r$ . For an even $r$, as the time evolves, the necessary condition for guaranteeing the decay of the perturbation is $\Lambda B_2>b_1$, meaning that the parameter $B_1$ can be  negative, or positive. On the other hand, for an odd $r$, the requirement for getting the decay of the perturbation is $\Lambda B_2<B_1$, meaning that $B_1>0$.\par

$\bullet$ {\bf Treating the model $f(T)=\frac{A_1 T^n+\Lambda}{A_2T^n+1}$}.\par  
Here, $n<0$, and the integral can be performed as
\begin{eqnarray}
-\frac{1}{2}\int C_bdt=-\frac{3}{4a_0}\Big[\ln{(t)}-\frac{(2n+3)(A_1-\Lambda A_2)t^2}{2(A_2+t^{2n})}+\frac{n(A_1-\Lambda A_2)t^2}{(A_2+t^{2n})^2}\nonumber\\
+\frac{3t^2(A_1-\Lambda A_2)}{2A_2}\,_2F_1\left(\frac{1}{n},1,1+\frac{1}{n};-\frac{t^{2n}}{A_2}\right)\Big]\,\,\,, \label{josiane}
\end{eqnarray}
with 
\begin{eqnarray}
C_b=\frac{3}{2a_0}\Big[\frac{n(2n+1)(A_1-\Lambda A_2)t^{-2n+1}}{(A_2t^{-2n}+1)^2}-\frac{4A_2n^2(A_1-\Lambda A_2)t^{-4n+1}}{(A_2t^{-2n}+1)^3}\Big]\,\,.
\end{eqnarray}
As  the time evolves, the argument of the hypergeometric function tends to zero and the hypergeometric function tends to $1$. Thus, the dominant term in (\ref{josiane})
reads
\begin{eqnarray}
\frac{3}{4a_0A^2_2}(A_2-n)(A_1-\Lambda A_2)t^2\,\,\,.\label{josiane2}
\end{eqnarray}
Here, one can distinguish two cases: ($A_2-n> 0$ and $A_1<\Lambda A_2$)  and ($A_2-n<0$ and $A_1>\Lambda A_2$). In the first case, one  gets $A_2>n$ meaning  that $A_2$ can be positive, or negative but with $|A_2|<|n|$. When $A_2>0$, $A_1$ can be  positive or negative, due to the relation $A_1< \Lambda A_2$, while for $A_2<0$, $A_1$ is necessarily negative. In the second case, one gets $A_2<n$, meaning that $A_2<0$, which allows $A_1$ to be positive, due to the relation $A_1>\Lambda A_2$.\par
We observe that some of the conditions for which the stability occurs, are also compatible with some energy conditions. This shows that for some values of the input parameters, acceptable models can be obtained, at least regarding to the energy conditions, the stability, the late time acceleration of the universe and the avoidance of the Big Rip.\par  
$\bullet$ {\bf Treating the model $f(T)=a_3\ln^q{\left(b_3 T^m\right)}$}.\par
As we have done  in the previous cases, the integral can be performed, yielding
\begin{eqnarray}
-\frac{1}{2}\int C_b =-\frac{3}{4a_0}\Big[\ln{(t)}+\frac{1}{2}a_3qb_3^{\frac{1}{m}}m^q\Gamma[q,\frac{g(t)}{m}]+a_3q(q-1)m^qb_3^{\frac{1}{m}}\Gamma[q-1,\frac{g(t)}{m}]\Big]\,\,\,, g(t)=\ln{(b_3t^{-2m})}\,\,\,.
\end{eqnarray}
with 
\begin{eqnarray}
C_b=\frac{3}{2a_0}\Big[\frac{1}{t}+a_3qm\, t\ln^{q-1}{(b_3t^{-2m})}+2a_3q(q-1)m^2t\ln^{q-2}{(b_3t^{-2m})}\Big]\,\,\,.
\end{eqnarray}
As we have previously shown, this model cures the Big Rip for $q>2$ and $m<0$. With these conditions, as the time evolves, only  the term $\ln{(t)}$ grows. Since $-3\ln{(t)}/(4a_0)$ is negative for large value of the time, it is easy to observe that the perturbation decays, and this corresponds to the  stability of the power-law solutions with this model. Observe that in this case, the constraints on the parameters $q$ and $m$ for which all the energy conditions are satisfied, leads to the stability of the power-law solutions. Thus, regarding to the stability, the energy conditions, the late time acceleration of the universe and the avoidance of the Big Rip, we can conclude that this model can be cosmologically acceptable for $a_3\neq 0$, $b_3>0$, $q>2$ and $m<0$.

\section{Discussions}
In this work we discussed the viability of an interesting alternative gravitational theory, namely, modified $f(R, T)$ theory of gravity, where $R$ is the curvature scalar and $T$ the trace of the energy momentum tensor. For our purpose we took the case where the algebraic function f(R, T) is cast into $f(R, T)= f_1(R)+f_2(T)$, and focused our attention to the assumption $f(R)=R$ and $f_2(T)=2f(T)$. The aim of the work is to analyse the viability of two models, $f(T)=(a_1T^n+b_1)/(a_2T^n+b_2)$ and $f(T)=a_3\ln^{q}{\left(b_3T^m \right)}$, (where $a_1$, $a_2$, $b_1$, $b_2$, $n$, $a_3$, $b_3$, $q$ and $m$ are input parameters) by investigating for which range of each parameters, the models  satisfy all the energy conditions or some of them. We also assumed that the usual matter content of the universe is pressureless (dust), for which the trace $T=\rho$. Depending of having $a_2\neq 0$ or $b_2\neq 0$, and according to the fact that the present stage of the universe may be characterized by the $\Lambda$CDM model, we distinguished two sub-models for the first one. Thus for $a_2\neq 0$ and $n>0$, the first model reduces into $f(\rho)=(\Lambda \rho^n+B_1)/(\rho^n+B_2)$, where $B_1=b_1/a_2$ and $B_2=b_2/a_2$, while for $b_2\neq 0$ and $n<0$, the model reduces into $f(\rho)=(A_1\rho^n+\Lambda)/(A_2\rho^n+\Lambda)$, where $A_1=a_1/b_2$ and $A_2=a_2/b_2$.  Energy conditions are generalised to the $f(R, T)$ by acting directly on the effective energy density and pressure, respectively $\rho_{eff}$ and $p_{eff}$ of the corresponding dark fluid. The same energy conditions, the null, weak, strong and dominant, applied to the ordinary energy density and pressure in GR  are then valid for the effective energy density and pressure. Besides to the evident constraints on the input parameters for each of the  expressions of $f(T)$, for obtaining  $R+2f(T)$ gravity models that satisfy energy conditions, we also presented figures characterizing models satisfying energy conditions, and whose parameters cannot be determined analytically, but rather, numerically. For plotting the graph, the current  value of the cosmological constant is used $\Lambda \sim1.7\times 10^{-121}$ Planck units \cite{barrow}, and the value of the energy density of the usual matter is considered as $\rho\sim 0.1\times 10^{-121}$ Planck units. The input parameters are normalized to $10^{-121}$ Plank units, such the cosmological constant and the energy density used for plotting the graph are $1.7$ and $0.1$, respectively. We  also discussed the situations where all or some of the energy conditions are satisfied. We observe that all the conditions which lead to the accomplishment of the NEC also satisfy the WEC, the SEC and the DEC. This is an obvious deduction from (\ref{vincent14})-(\ref{vincent17}). In the same way, it is easy to observe that all the conditions for  which the WEC is satisfied, also lead to the accomplishment of the DEC. From the same expressions (\ref{vincent15})-(\ref{vincent16}), we see that the sufficient condition for the requirement of satisfying the WEC leads to the accomplishment of the SEC is $f(T)< 0$. Thus, among the conditions for which the WEC is satisfied, those for which the function is always negative, also lead to the accomplishment of the SEC. However, about the SEC with respect to the DEC, the situation is quite different. The requirement for the conditions that satisfy the SEC leading to the accomplishment of the DEC, is always having  $f(T)>0$. This means that, among the conditions for which the SEC is satisfied, those for which the function $f(T)$ is always positive, guarantee the accomplishment of the DEC  \par
Moreover, in order to check the viability of the models in some ways, we investigate the stability of de Sitter solutions and power-law solutions under perturbations in the FRW metric in the framework of $R+2f(T)$ gravity, by using the models proposed in this work. The results show that only for the models $f(T)=(\Lambda T^n+B_1)/(T^n+B_2)$ and $f(T)=a_3\ln^{q}{\left(b_3T^m\right)}$  de Sitter solutions can present stability.  However, for the power-law solutions, the stability can be obtained for each model under some conditions. We observe that some of the conditions that lead to the stabilities of de Sitter solutions and power-law solutions are compatible with the accomplishment of some of the energy conditions, the late-time cosmic acceleration and the avoidance the type-I future finite time singularity (Big Rip). We conclude that, in the framework of $R+2f(T)$, viable models of type $f(T)=\frac{a_1T^n+b_1}{a_2T^n+b_2}$ and $f(T)=a_3\ln^{q}{\left(b_3T^m\right)}$ can be obtained, when the usual matter is assumed as pressureless.\par
Despite to these satisfactory results obtained in the framework of $R+2f(T)$ gravity, it is important to recall that this work may be view as a particular case of  the general one where $f(R, T)= f_1(R)+f_2(T)$, with $f_1(R)\neq R$. Our purpose for a coming work is to undertake this general case, for which the $f(R)$ gravity is also a particular case, and analyse the energy conditions,  the stability of some known scale factor solutions, against homogeneous perturbations and also the cosmological background evolution.    

\par

\vspace{1cm}

{\bf Acknowledgement:}  M. J. S. Houndjo thanks prof. S.D.Odintsov for useful suggestions and also CNPq/FAPES for financial support. A. V. Monwanou thanks IMSP-UAC  for financial support. The authors also thank very much the referees for useful suggestions for the reorganization of the manuscript.


\newpage
\begin{figure}[t]
\begin{minipage}[t]{0.45\linewidth}
\includegraphics[width=\linewidth]{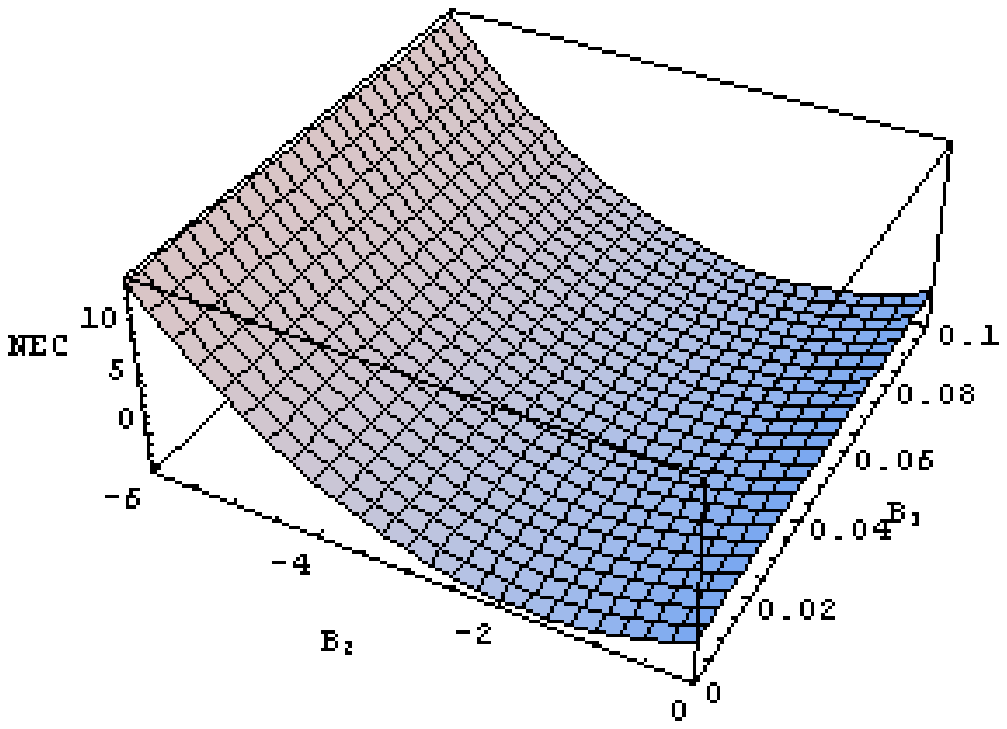}
\end{minipage} \hfill
\begin{minipage}[t]{0.45\linewidth}
\includegraphics[width=\linewidth]{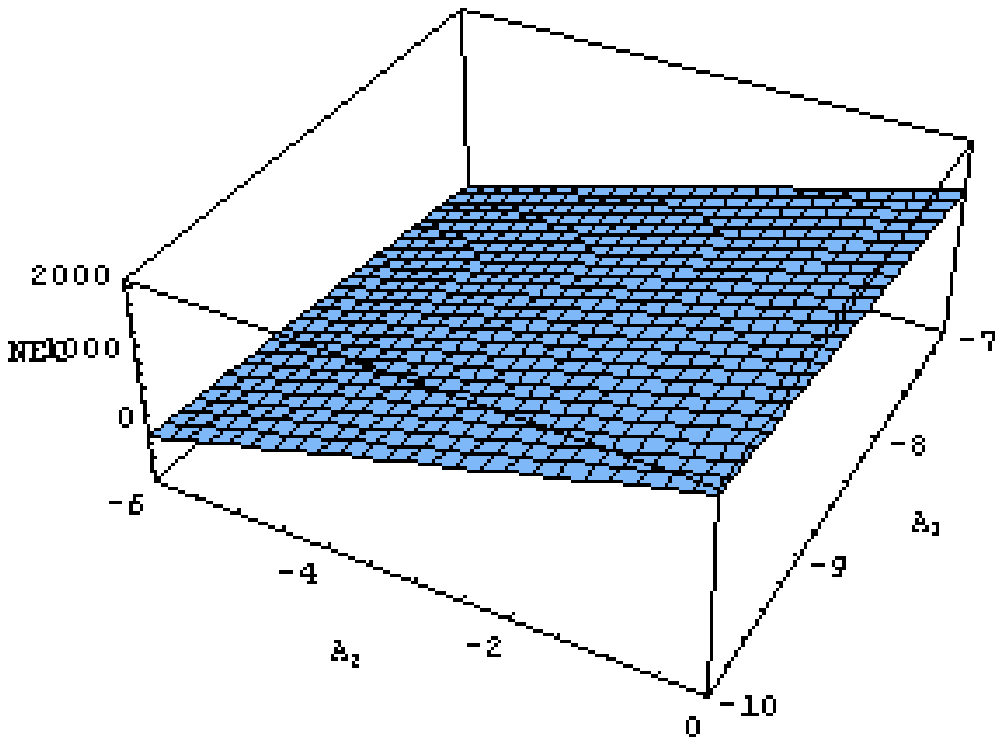}
\end{minipage}\hfill
\caption{{\protect\footnotesize The graph at the left side of this figure corresponds to the NEC by plotting  $2n\rho^{n-1}(\Lambda B_2-B_1)+(\rho^n+B_2)^2$  in functions of $B_1$ and $B_2$ with $\rho=0.1$, $n=1$,  $\Lambda=1.7$.  This case is different from the evident conditions. Here, $2n\rho^{n-1}(\Lambda B_2-B_1)$, but the NEC is realized. 
The graph at the right side of the figure corresponds to the NEC but in this case by plotting $2n\rho^{n-1}(A_1-A_2\Lambda)+(A_2\rho^n+1)^2$ in functions of $A_1$ and $A_2$ with $n=-1$,  $\Lambda=1.7$ an $\rho=0.1$. This situation  is also different from those of the evident conditions, but also  realizes the NEC.}}
\label{}
\end{figure}

\begin{figure}[t]
\begin{minipage}[t]{0.45\linewidth}
\includegraphics[width=\linewidth]{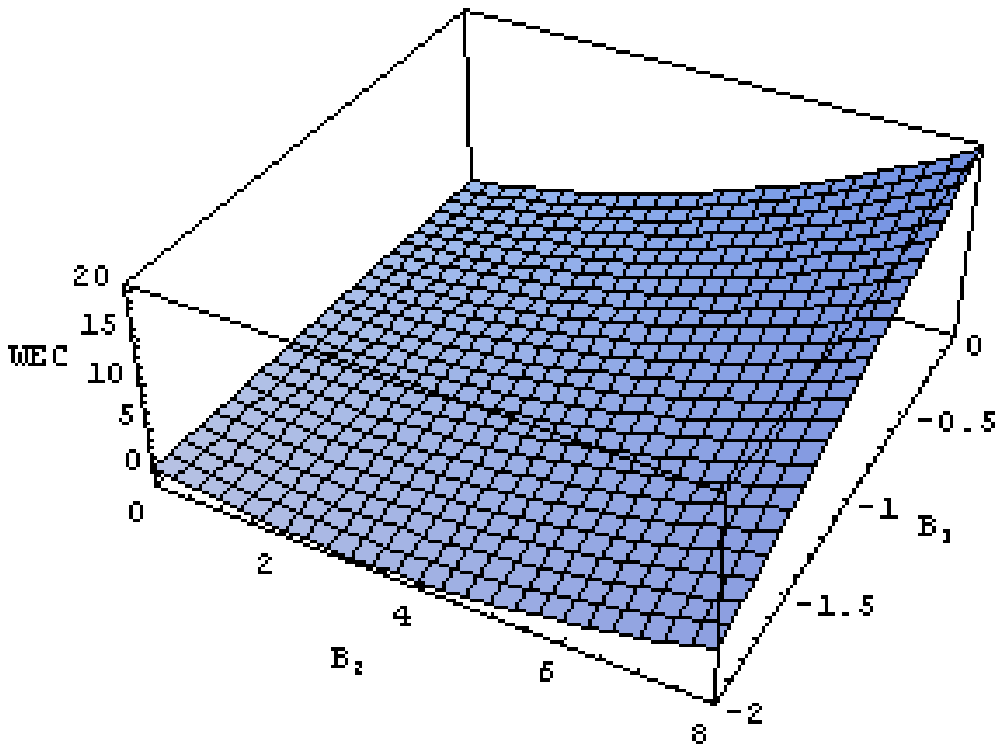}
\end{minipage} \hfill
\begin{minipage}[t]{0.45\linewidth}
\includegraphics[width=\linewidth]{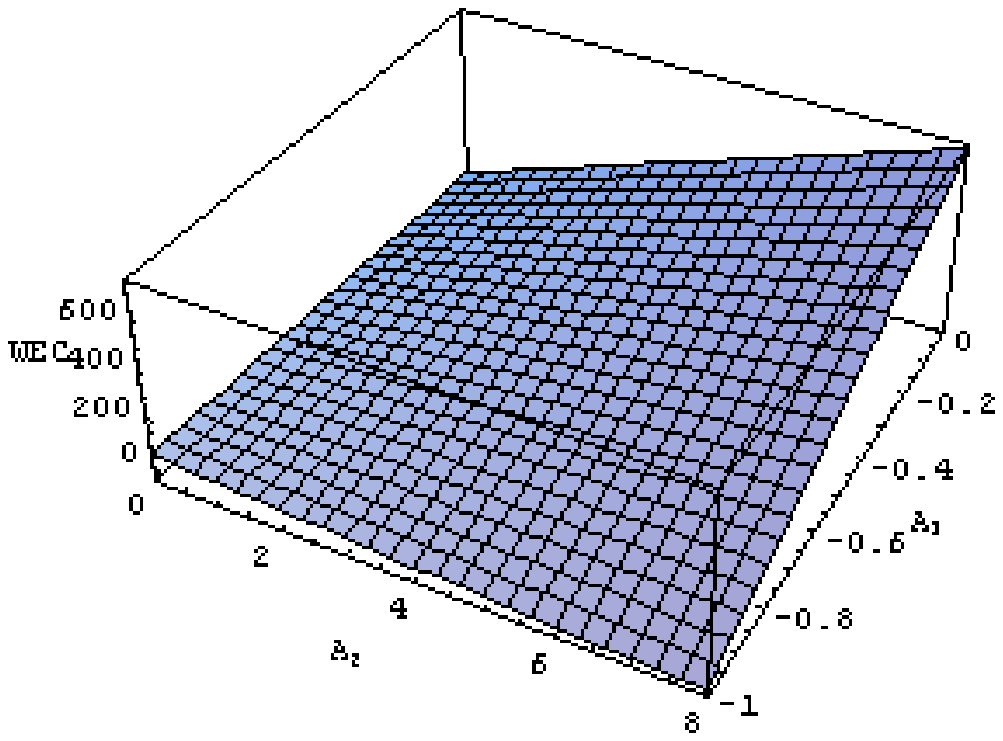}
\end{minipage}\hfill
\caption{{\protect\footnotesize The graph at the left side is obtained by plotting the function $\Lambda\rho^{2n}+2\rho^{2n+1}+4B_2 \rho^{n+1}+2B^2_2\rho+\left[(1+4n)\Lambda B_2+(1-4n)B_1\right]\rho^n+B_1B_2$ in terms of suitable values of $B_1$ and $B_2$, with 
$n=1$, $\Lambda=1.7$ and $\rho=0.1$. We have a situation different from the evident ones but the WEC is satisfied. At the right side of the figure one has the graph obtained by plotting $
A_1A_2\rho^{2n}+2\rho^{2n+1}+4A_2 \rho^{n+1}+2\rho+\left[(1+4n)A_1+(1-4n)A_2\Lambda\right]\rho^n+\Lambda $ in terms of $A_1$ and $A_2$, with $n=-1$, $\Lambda=1.7$ and $\rho=0.1$. We also have here a case different from the evident ones but for which the WEC is still  realized.}}
\label{}
\end{figure}

\begin{figure}[t]
\begin{minipage}[t]{0.45\linewidth}
\includegraphics[width=\linewidth]{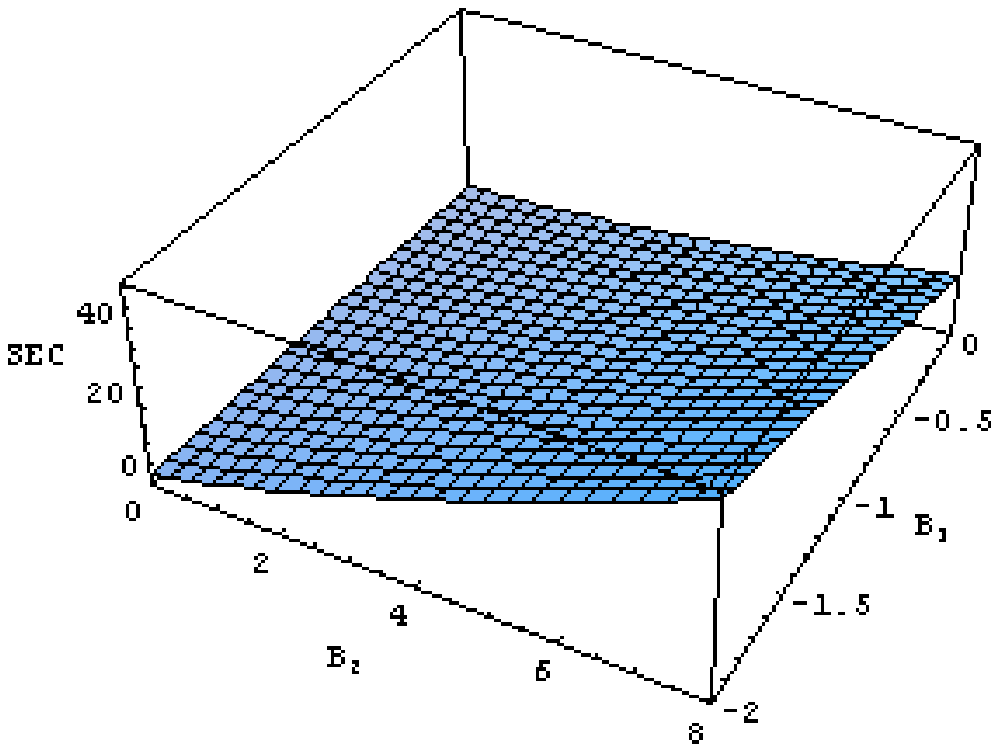}
\end{minipage} \hfill
\begin{minipage}[t]{0.45\linewidth}
\includegraphics[width=\linewidth]{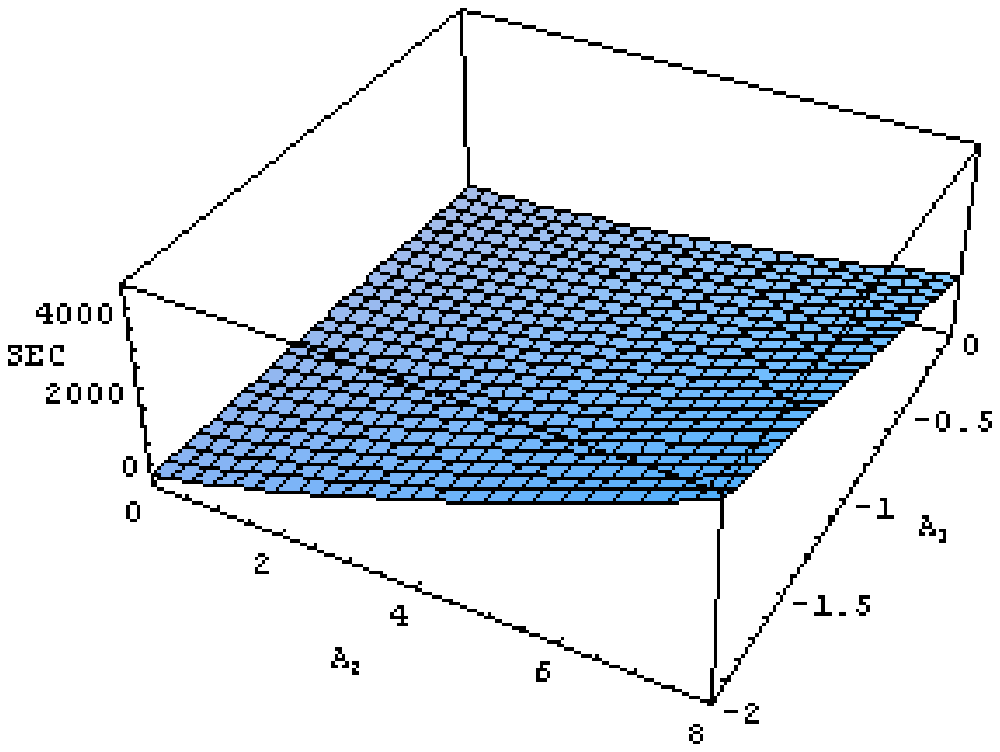}
\end{minipage}\hfill
\caption{{\protect\footnotesize At the left side the SEC is represented by plotting $2\rho^{2n+1}-2\Lambda\rho^{2n}+4B_2\rho^{n+1}+2B^2_2\rho-
2\left[(1-2n)\Lambda B_2+(1+2n)B_1\right]\rho^n-2B_1B_2 $ in functions of $B_1$ and $B_2$, setting  $\Lambda=1.7$, $\rho=0.1$ and $n=1$.  The graph at right side of the figure represents the  SEC where $
2A_2^2\rho^{2n+1}-2A_1A_2\rho^{2n}+4A_2\rho^{n+1}+2\rho-
2\left[(1-2n)A_1+(1+2n)A_2\Lambda \right]\rho^n-2\Lambda$ is plotted in terms of  $A_1$ and $A_2$, setting $n=-1$, $\rho=0.1$ and $\Lambda=1.7$.}}
\label{}
\end{figure}

\begin{figure}[t]
\begin{minipage}[t]{0.45\linewidth}
\includegraphics[width=\linewidth]{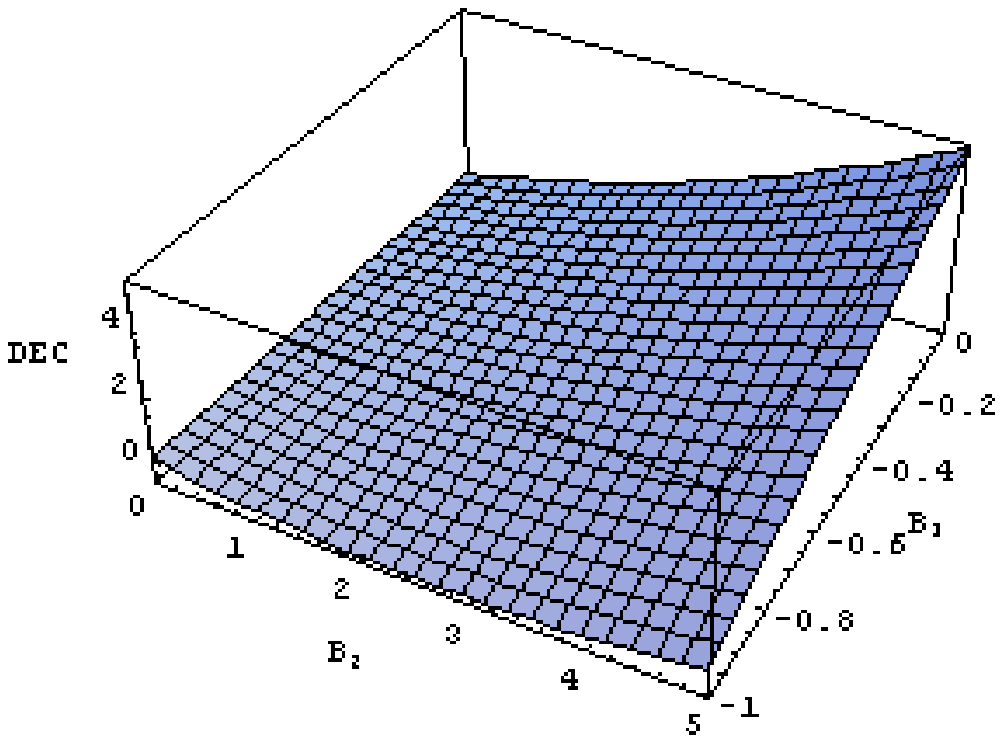}
\end{minipage} \hfill
\begin{minipage}[t]{0.45\linewidth}
\includegraphics[width=\linewidth]{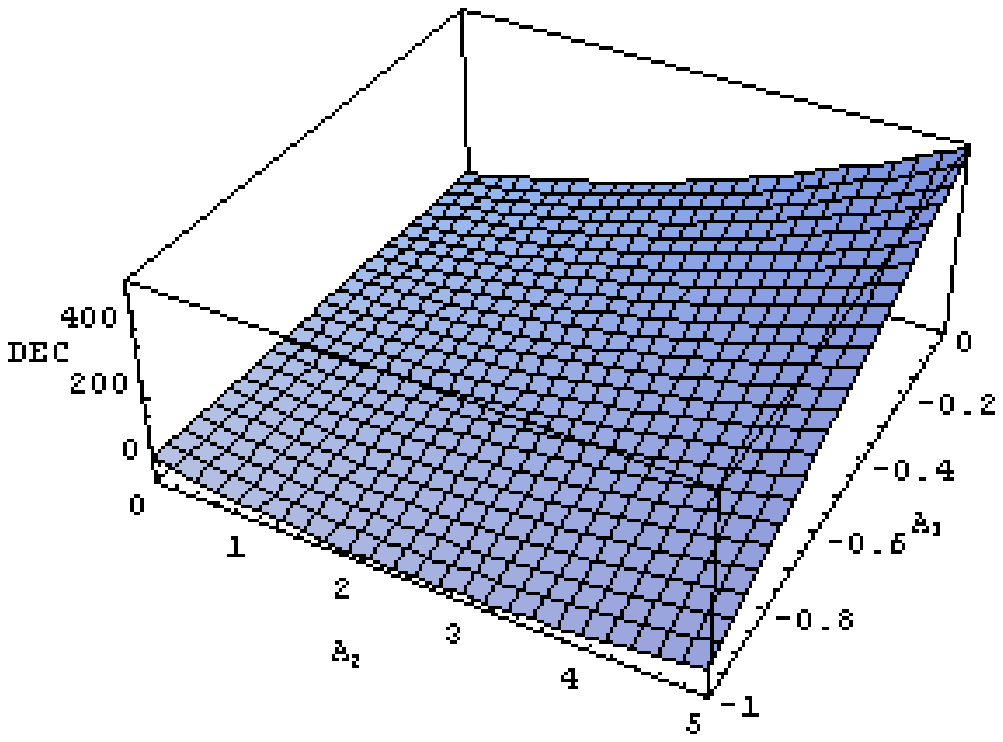}
\end{minipage}\hfill
\caption{{\protect\footnotesize The graph at the left side of this figure characterizes the DEC, coming from $\rho^{2n+1}+\Lambda\rho^{2n}+2B_2\rho^{n+1}+B_2^2\rho+B_1B_2
+\left[(2n+1)\Lambda B_2+(1-2n)B_1\right]\rho^n $ plotted in functions of $B_1$ and $B_2$,  by
setting $n=1$, $\rho=0.1$  and $\Lambda=1.7$, while the graph at the right side represent the DEC from the plot of $
A^2_2\rho^{2n+1}+A_1A_2\rho^{2n}+2A_2\rho^{n+1}+\rho+\Lambda 
+\left[(2n+1)A_1+(1-2n)\Lambda A_2\right]\rho^n$ in functions of $A_1$ and $A_2$, with $n=-1$, $\rho=0.1$ and $\Lambda=1.7$. }}
\label{}
\end{figure}

\begin{figure}[t]
\begin{minipage}[t]{0.45\linewidth}
\includegraphics[width=\linewidth]{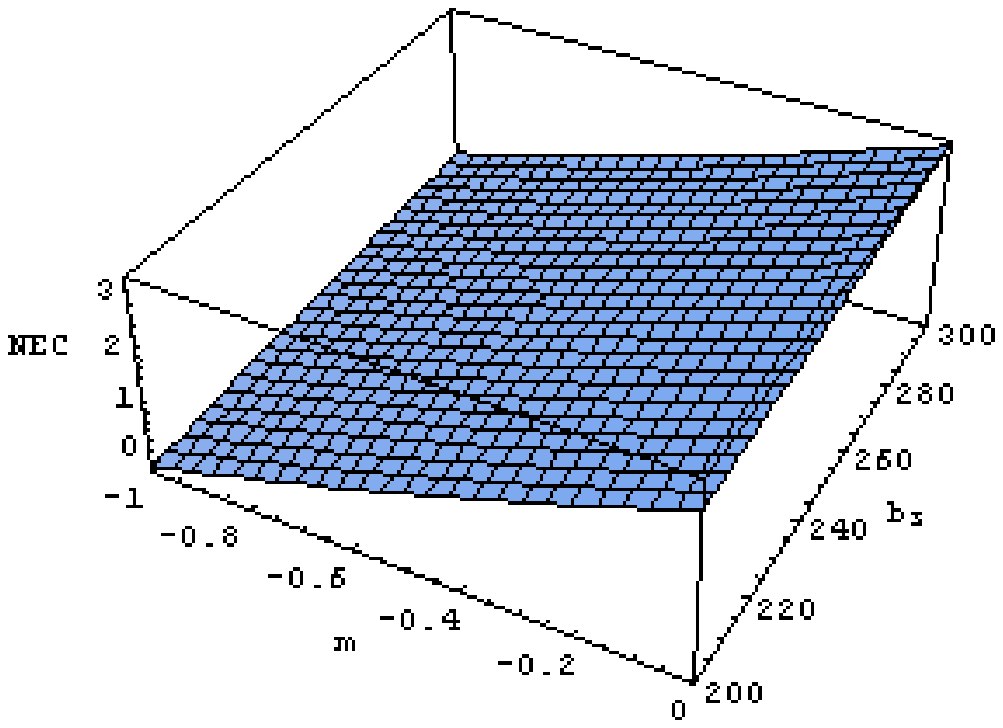}
\end{minipage} \hfill
\begin{minipage}[t]{0.45\linewidth}
\includegraphics[width=\linewidth]{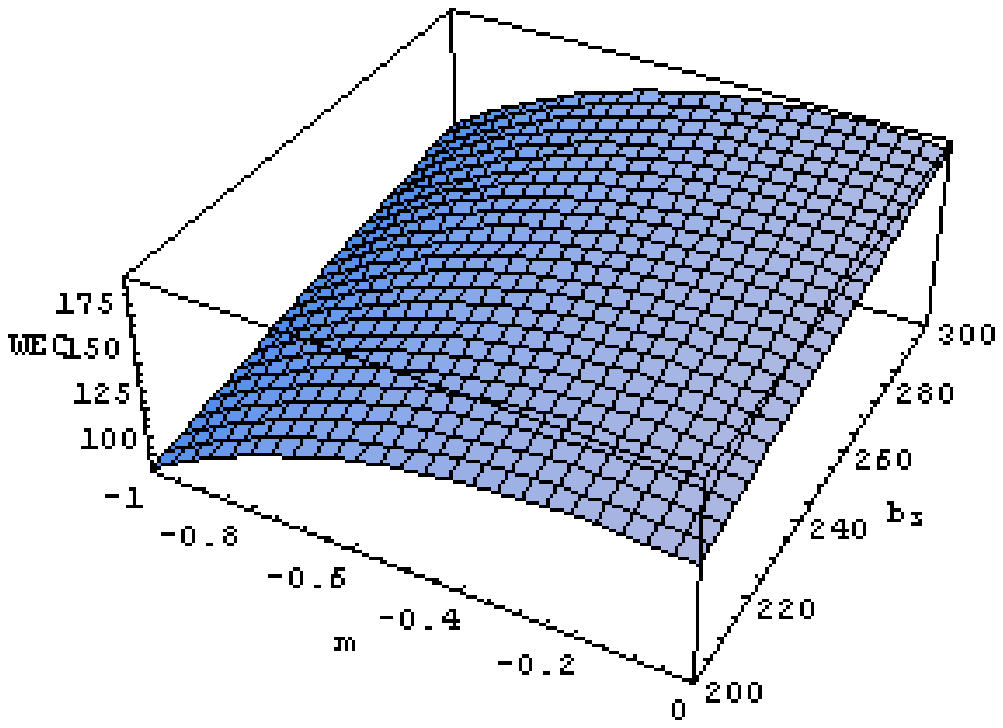}
\end{minipage}\hfill
\caption{{\protect\footnotesize The graph at the left side of this figure represents the NEC and is obtained by plotting $\rho+2q m a_3+\rho\ln^{q-1}{\left(b_3\rho^m\right)}$ in functions of $m$ and $b_3$, setting  $\rho=0.1$, $a_3=1$  and  $q=3$. This case is different from the evident conditions but still realizes the NEC.  The graph at the  right side of the figure corresponds to the WEC, plotting $2\rho+2 q m a_3+ \left(\rho+2qma_3 \right)\ln^{q-1}{\left(b_3\rho^m\right)}+a_3\ln^{q}{\left(b_3\rho^m\right)}$ in functions of $m$ and $b_3$, where we set $q=3$, $a_3=1$ and $\rho=0.1$.}}
\label{}
\end{figure}

\begin{figure}[t]
\begin{minipage}[t]{0.45\linewidth}
\includegraphics[width=\linewidth]{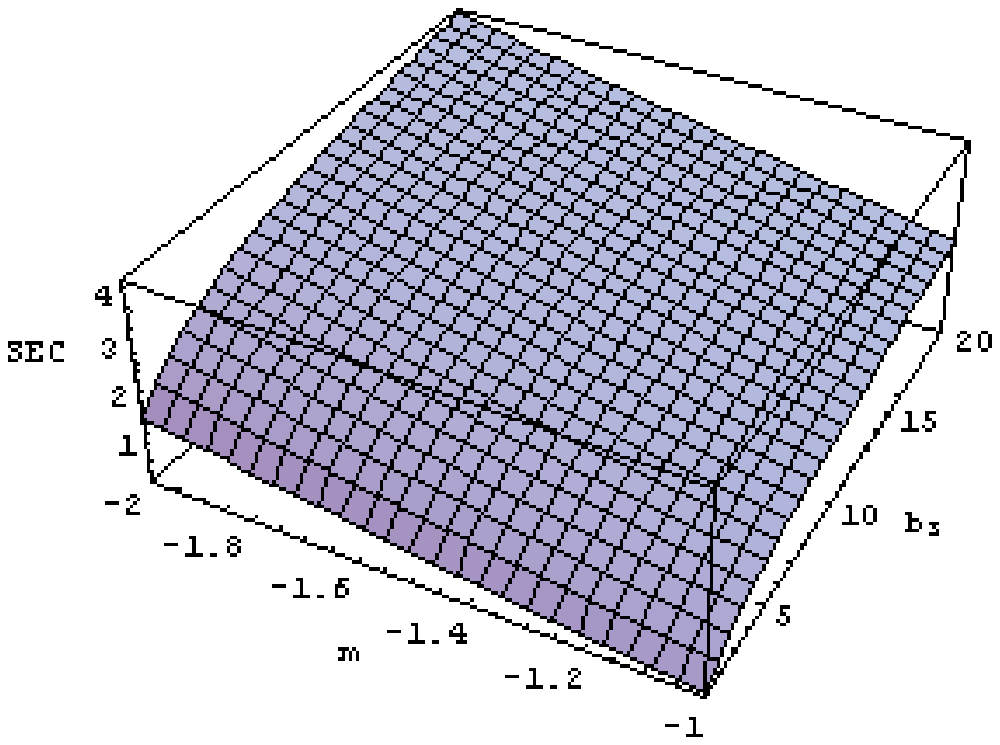}
\end{minipage} \hfill
\begin{minipage}[t]{0.45\linewidth}
\includegraphics[width=\linewidth]{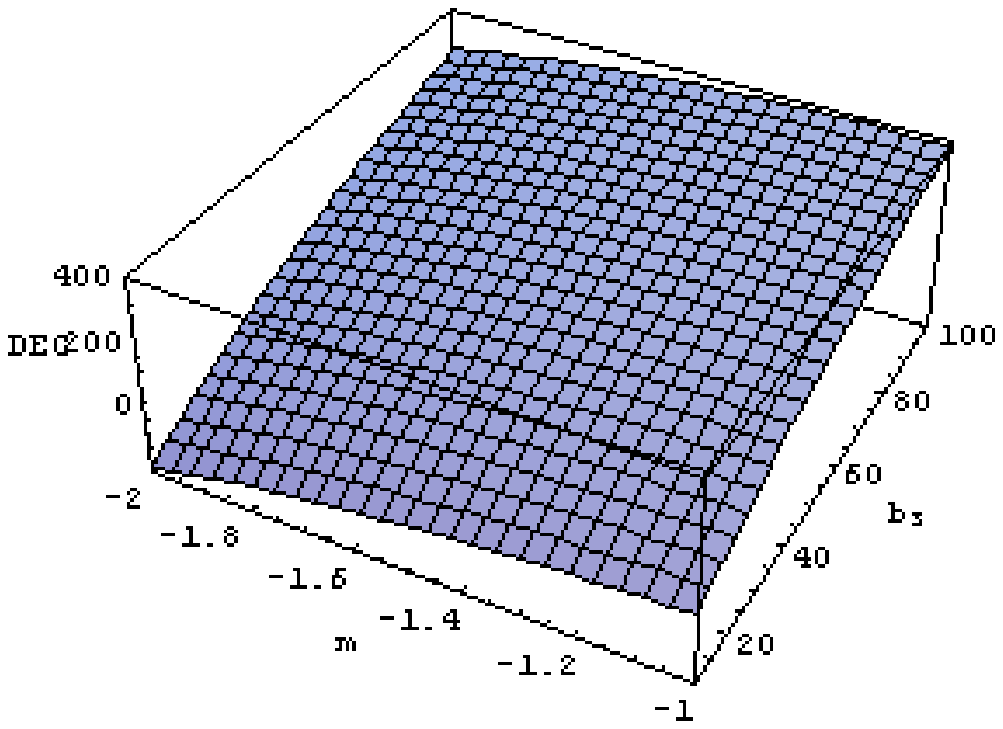}
\end{minipage}\hfill
\caption{{\protect\footnotesize The graph at the left side characterizes the SEC is obtained by plotting $2\rho+2qma_3+\left( \rho+2qma_3\right)\ln^{q-1}{\left(b_3\rho^m\right)}-2a_3\ln^{q}{\left(b_3\rho^m\right)}$ in functions of $m$ and $b_3$ where we set $\rho=0.1$, $a_3=10^{-3}$ and $q=3$. This situation is different from the evident conditions, but  also realizes the SEC. The graph at the right side of the figure is obtained by plotting $3\rho+2qma_3+\left(\rho+4qma_3\right)\ln^{q-1}{\left(b_3\rho^m\right)}+3a_3\ln^{q}{\left(b_3\rho^m\right)}$ in functions of $m$ and $b_3$, with $q=3$,  $a_3=1$ and $\rho=0.1$. This graph represents the DEC.}}
\label{}
\end{figure}



\end{document}